\documentclass[manuscript, nonacm]{acmart}

\AtBeginDocument{%
  \providecommand\BibTeX{{%
    \normalfont B\kern-0.5em{\scshape i\kern-0.25em b}\kern-0.8em\TeX}}}

\acmConference[Conference abbreviation]{Full conference name}{Date}{Location}

\usepackage{type1cm} % type1 computer modern font
\usepackage{graphicx} % advanced figures
\usepackage{xspace} % fix space in macros
\usepackage{balance} % to better equalize the last page
\usepackage{booktabs} % nicer tables
\usepackage{multirow} % multi rows for tables
\usepackage[font={bf}, tableposition=top]{caption} % captions on top for tables
\usepackage{subcaption} % subfloats
\usepackage{bold-extra} % bold + {small capital, italic}
\usepackage{microtype} % compress text
\usepackage{siunitx} % \num for decimal grouping
\usepackage{xfrac} % nicer slanted fractions
\usepackage{mathtools} % amsmath++
\PassOptionsToPackage{hyphens}{url} % acmart compatibility
\PassOptionsToPackage{bookmarks, pdftex, colorlinks=true, pagebackref=true, backref=page}{hyperref} % acmart compatibility
\usepackage{cleveref} % smart references
    \crefname{table}{Tbl.}{Tbls.}
    \Crefname{table}{Tbl.}{Tbls.}
    \crefname{figure}{Fig.}{Figs.}
    \Crefname{figure}{Fig.}{Figs.}
    \Crefname{appendix}{Supplementary Material}{Supplements}

\PassOptionsToPackage{square,numbers}{natbib} % acmart compatibility
\usepackage{hyphenat} % name in single line
\usepackage{ragged2e}
\newenvironment{squishlist}
{\begin{list}{$\bullet$}
 {\setlength{\itemsep}{0pt}
     \setlength{\parsep}{3pt}
     \setlength{\topsep}{3pt}
     \setlength{\partopsep}{0pt}
     \setlength{\leftmargin}{1.5em}
     \setlength{\labelwidth}{1em}
     \setlength{\labelsep}{0.5em} } }
{\end{list}}

\usepackage{xcolor}
\hypersetup{
   colorlinks=true,
   linkcolor={blue},
   filecolor={green!50!black},
   citecolor={black!50!black}, 
   urlcolor={blue!80!black},
}

\newcommand{\greenparty}{\emph{Gr\"une}\xspace}

\definecolor{linkecolor}{HTML}{990099}
\definecolor{greenscolor}{HTML}{46962B}
\definecolor{spdcolor}{HTML}{E3000F}
\definecolor{fdpcolor}{HTML}{DFD220}
\definecolor{unioncolor}{HTML}{000000}
\definecolor{afdcolor}{HTML}{009EE0}

\newcommand{\figletter}[1]{{{\fontfamily{\sfdefault}\selectfont \textbf{#1}}}}

\usepackage{tikz}

\newcommand{\tikzcircle}[2][red,fill=red]{\tikz[baseline=-0.5ex]\draw[#1,radius=#2] (0,0) circle ;}%

\begin{document}

\title{Systematic discrepancies in the delivery of political ads on Facebook and Instagram}

\author{Dominik Bär}
\authornote{Both authors contributed equally to this research.}
\authornote{To whom correspondence should be addressed. E-mail: \href{mailto:baer@lmu.de}{baer@lmu.de}; Mail: LMU Munich, Geschwister-Scholl-Platz 1, 80539 München}
\affiliation{%
  \institution{LMU Munich,}
  \institution{Munich Center for Machine Learning}
  \city{Munich}
  \country{Munich, Germany}
}

\author{Francesco Pierri}
\authornotemark[1]
\affiliation{%
 \institution{Politecnico di Milano,}
  \institution{Dipartimento di Elettronica, Informazione e Bioingegneria}
 \city{Milan}
 \country{Milan, Italy}
 \email{francesco.pierri@polimi.it}}

\author{Gianmarco De Francisci Morales}
\affiliation{%
 \institution{CENTAI}
 \city{Turin}
 \country{Turin, Italy}
 \email{gdfm@acm.org}}

\author{Stefan Feuerriegel}
\affiliation{%
  \institution{LMU Munich,}
  \institution{Munich Center for Machine Learning}
  \city{Munich}
  \country{Munich, Germany}
  \email{feuerriegel@lmu.de}}

\renewcommand{\shortauthors}{}

\begin{abstract}
\textbf{Abstract:} Political advertising on social media has become a central element in election campaigns. However, granular information about political advertising on social media was previously unavailable, thus raising concerns regarding fairness, accountability, and transparency in the electoral process. In this paper, we analyze targeted political advertising on social media via a unique, large-scale dataset of over \num{80000} political ads from Meta during the 2021 German federal election, with more than $1.1$ billion impressions. For each political ad, our dataset records granular information about targeting strategies, spending, and actual impressions. We then study ($i$)~the prevalence of targeted ads across the political spectrum; ($ii$)~the discrepancies between targeted and actual audiences due to algorithmic ad delivery; and ($iii$)~which targeting strategies on social media attain a wide reach at low cost. We find that targeted ads are prevalent across the entire political spectrum. Moreover, there are considerable discrepancies between targeted and actual audiences, and systematic differences in the reach of political ads (in impressions-per-EUR) among parties, where the algorithm favor ads from populists over others.

\vspace{0.2cm}
\noindent\textbf{Significance:} Social media platforms have become important tools for political campaigning worldwide. In our study of over \num{80}k political ads from the 2021 German federal election, we reveal extensive use of targeted political advertising across the full political spectrum, significant discrepancies between targeted and actual audiences, and a systematic bias in the algorithmic delivery of ads favoring populist parties. These findings highlight the complex relationship between digital campaigning strategies and democratic processes, and caution about the potential for algorithmic biases to influence election campaigns. Overall, our work contributes to a better understanding of targeted political advertising on social media and informs policymakers about the design of effective regulatory frameworks to promote fairness, accountability, and transparency.
\end{abstract}


\maketitle

\newpage
\section*{Introduction}

With around 4.6~billion users globally~\cite{Statista.2023}, social media platforms such as Facebook, Instagram, and Twitter/X have become important tools for political campaigning worldwide \cite{Votta.2024}. For example, in the U.S., expenditure on online political advertising rose from USD~$\sim$70~million in 2014 to USD~$\sim$1.8~billion in 2018~\cite{Fowler.2020}. Similarly, politicians in Europe have recognized the importance of social media for their campaigns. For instance, the majority of candidates in the 2021 German federal election believed that social media can influence voters~\cite{GLES.2022}.

An essential feature of advertising on social media is \emph{targeting}, which allows advertisers to select specific user groups and deliver tailored political messages to particularly receptive audiences \cite{Matz.2017,Breza.2021,Goldberg.2021, Votta.2024}. For example, campaigns can send tailored ads that align with the interests of distinct voter groups~\cite{Ridout.2021}, thus ensuring that their content resonates with the unique political perspectives of each audience~\cite{Fowler.2021}. However, targeting in political advertising is problematic~\cite{Dommett.2019c, Imana.2021, Tappin.2023}. First, targeting is concerning if parties cater ads to specific groups~\cite{Hersh.2013} or send conflicting messages on political issues to different audiences \cite{Votta.2023}. Second, targeted ads are distributed by proprietary algorithms that are beyond societal scrutiny and that may exhibit biases that influences the audience of specific ads~\cite{Lambrecht.2019, Ali.2019, Ali.2021, Imana.2021}. Third, privacy concerns are eminent given that political targeting heavily depends on potentially sensitive information (e.g., ethnic origin, sexual orientation) to identify receptive audiences~\cite{Korolova.2010, Auxier.2020, Cabanas.2021}. Essentially, the use of targeting raises concerns regarding fairness, accountability, and transparency in electoral processes.

The concerns regarding democratic integrity have spurred calls to monitor political advertising on social media~\cite{Isaak.2018, Dommett.2019b}. However, granular information about political ads (e.g., impression counts, price per ad) is in the hands of proprietary social media platforms and, so far, has been either unavailable or deemed imprecise~\cite{Edelson.2020, LePochat.2022}. This lack of transparency may prevent accountability for misconduct, which is particularly concerning given recent evidence suggesting that political advertising on social media directly affects voter turnout~\cite{Aggarwal.2023} and vote choice~\cite{Hager.2019, Coppock.2022, Bar.2024}. As such, there is a growing need to monitor targeted political advertising on social media to safeguard democratic integrity.

A combination of public pressure~\cite{Fowler.2020} and regulatory efforts (e.g., the Honest Ads Act in the U.S.~\cite{UnitedStatesHouseofRepresentatives.2021} and the Digital Services Act in the E.U.~\cite{EuropeanCommission.2022}) have pushed social media platforms to strengthen their transparency efforts around political advertising. Indeed, Meta has launched the Meta Ad Library, which provides public access to all political and social ads published on Facebook and Instagram, and allows researchers to study political advertising at scale~\cite{Capozzi.2020, Capozzi.2021, Capozzi.2023, Aisenpreis.2023, Pierri.2023} (see Supplementary Material 1 for a comprehensive overview of the literature). However, existing analyses had only limited access to political ads, since crucial information about \emph{targeting} was missing. As such, it remains unclear how targeting is used, how targeted ads are distributed, what the different targeting strategies behind political ads are, and how their reach at a given budget varies.

There are good reasons to believe that parties adopt diverse targeting strategies~\cite{Ridout.2021, Brodnax.2022, Fowler.2021}. Previous research on the content of online political ads has shown that parties pursue different communication strategies~\cite{Kruschinski.2022, Capozzi.2023}. Some parties advertise by particularly focusing on issues related to their base~\cite{Kruschinski.2022, Capozzi.2023} such as, for example, environmentalism in the case of the \greenparty. In contrast, other parties avoid ads related to specific political issues~\cite{Fowler.2021, Kruschinski.2022} and tend to publish more ``generic'' ads, for example, introducing a candidate or calling to vote. Since the latter does not refer to a concrete political issue, such ads may reach audiences across party boundaries, which may vary with respect to age, gender, and interests~\cite{Capozzi.2023}. As such, targeting may limit the political participation of disadvantaged groups due to a party's targeting strategy (differently from ads on broadcast media that can be received by all voter groups). Furthermore, there is reason to expect that the algorithmic delivery of social media ads may introduce further bias \cite{Lambrecht.2019, Ali.2019, Ali.2021}, thus resulting in differences between actual and intended audiences. For example, social media algorithms exhibit a tendency to target fewer women due to differences in advertising costs~\cite{Lambrecht.2019}. Such algorithmic bias can lead to discrimination as women may be less frequently exposed to political campaigns and thus harm political participation. Moreover, such bias can harm political competition when, for instance, some parties consistently pay higher prices for political ads, thus leading to fairness issues.

In this paper, we analyze targeted political advertising on social media using a large-scale dataset with $N=\num{81549}$ political ad contracts (henceforth simply ads)\footnote{Advertisers on Meta may purchase ads with the same creatives (i.e., content, image, etc.) multiple times. Therefore, we use `ad' to refer to the specific ad contract, including its timing, budget, and targeting settings besides its creative parts.} from Meta during the 2021 German federal election (see Supplementary Material 2 for additional context on the election). Overall, these ads generated more than 1.1 billion impressions with an overall cost of EUR~9.8 million. Our dataset provides a unique view of the targeting strategies that parties use across the entire political spectrum, during an election with more than $60$ million eligible voters. In particular, our dataset comprises granular information about each ad, including targeting strategies, spending, and actual impressions, and thus allows us to study targeted political advertising on social media.

Our analysis is three-fold: (\emph{i})~We assess the prevalence of targeted political ads across the full political spectrum and infer detailed targeting strategies used by political parties for their election campaigns. (\emph{ii})~We evaluate discrepancies between targeted and actual audiences due to algorithmic bias in the ad delivery and how such discrepancies vary across parties. (\emph{iii})~We analyze the characteristics of targeted ads with far reach at a given budget during elections, and analyze whether parties are discriminated by algorithmic ad delivery in that they pay a higher price per impression.

\section*{Results}  

\subsection*{Targeted political advertising during the 2021 German federal election}

We analyze targeted political ads on Meta during the 2021 German federal election (see Supplementary Material 2 for additional context on the election) and compare how parties across the political spectrum use targeting for their campaign purposes. Overall, we analyze $N=\num{81549}$ political ads that generated more than 1.1 billion impressions with an overall cost of EUR~9.8 million. For a breakdown of our dataset in terms of total number of ads, ad spending, and impressions by party see Supplementary Material 3. The ads in our dataset are designed mostly to mobilize and persuade voters and tend to focus on parties more broadly rather than candidates. Details are in Supplementary Material 4 and Supplementary Material 5.

Targeting has spurred concerns regarding political advertising on social media \cite{Dommett.2019, Tappin.2023}. However, it is unclear to what extent parties use targeting during election campaigns. Throughout the paper, we consider an ad to employ ``targeting'' if it uses any targeting category available to advertisers on Meta in addition to demographics (i.e., gender and age) and location (since the latter must always be specified in the ad creation). For details, see Section ``\nameref{sec:methods}.''

To analyze the prevalence of targeted political ads on Meta during the 2021 German federal election, we start by quantifying the number of targeted ads. In the run-up to the election, 72.3\,\% of all ads used targeting, which corresponds to 72.6\,\% of the total ad spending on Meta during the election. This highlights the importance of targeted ads for political campaigns on social media.

\begin{figure}[!t]
    \centering
    \includegraphics[width=.65\linewidth]{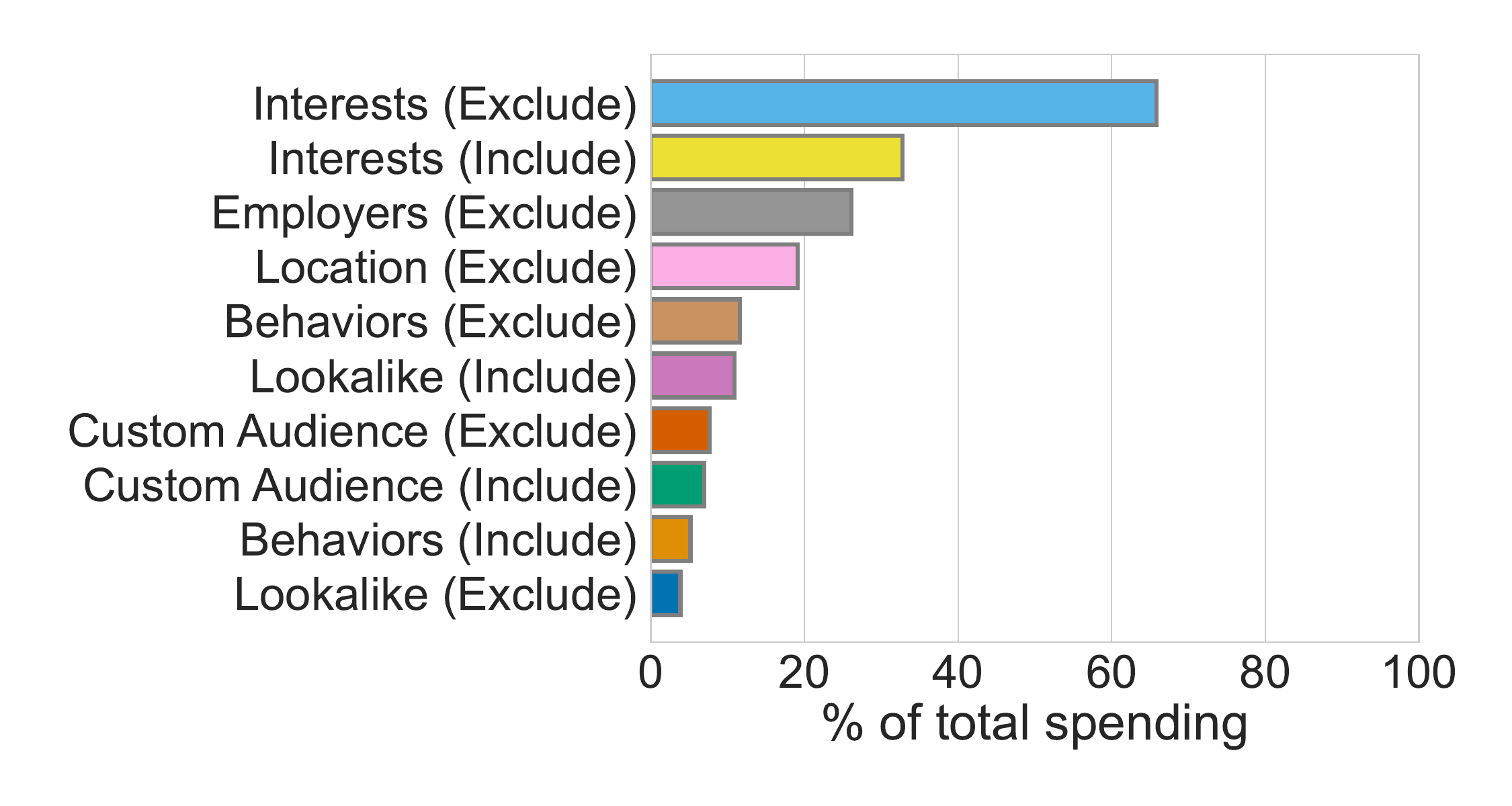}
    \caption{Top-10 targeting criteria by total spending.}
    \label{fig:prevalence_targeting_money}
\end{figure}

Meta allows advertisers to target users based on various targeting categories (see Supplementary Material 8 for an overview). We expect that some targeting categories are more popular than others and thus study how the campaign budget is distributed between categories. \Cref{fig:prevalence_targeting_money} shows the top-10 targeting categories in terms of spending across all parties. We find that parties tend to use exclusion rather than inclusion criteria to target users. This result suggests that most parties rely on broad audiences and allow Meta to optimize ad delivery among users. To define inclusion criteria, parties largely rely on so-called interests (e.g., social equality, environmentalism, and international relations), behaviors (e.g., early adopters of new technology, commuters, international travelers), or employers (e.g., business owners, police officers, Ford Deutschland). Parties also frequently define a specific list of users to be targeted (``Custom'') or an audience that is similar to a previously defined target group (``Lookalike''). Lastly, parties commonly target users based on their location. For example, 88.72\,\% of the ads have a precise location targeting (beyond Germany). Overall, these results show that parties employ a wide variety of targeting strategies. For a breakdown of the top-10 targeting criteria by party see Supplementary Material 3.

\begin{figure}[H]
    \centering
    \includegraphics[width=.65\linewidth]{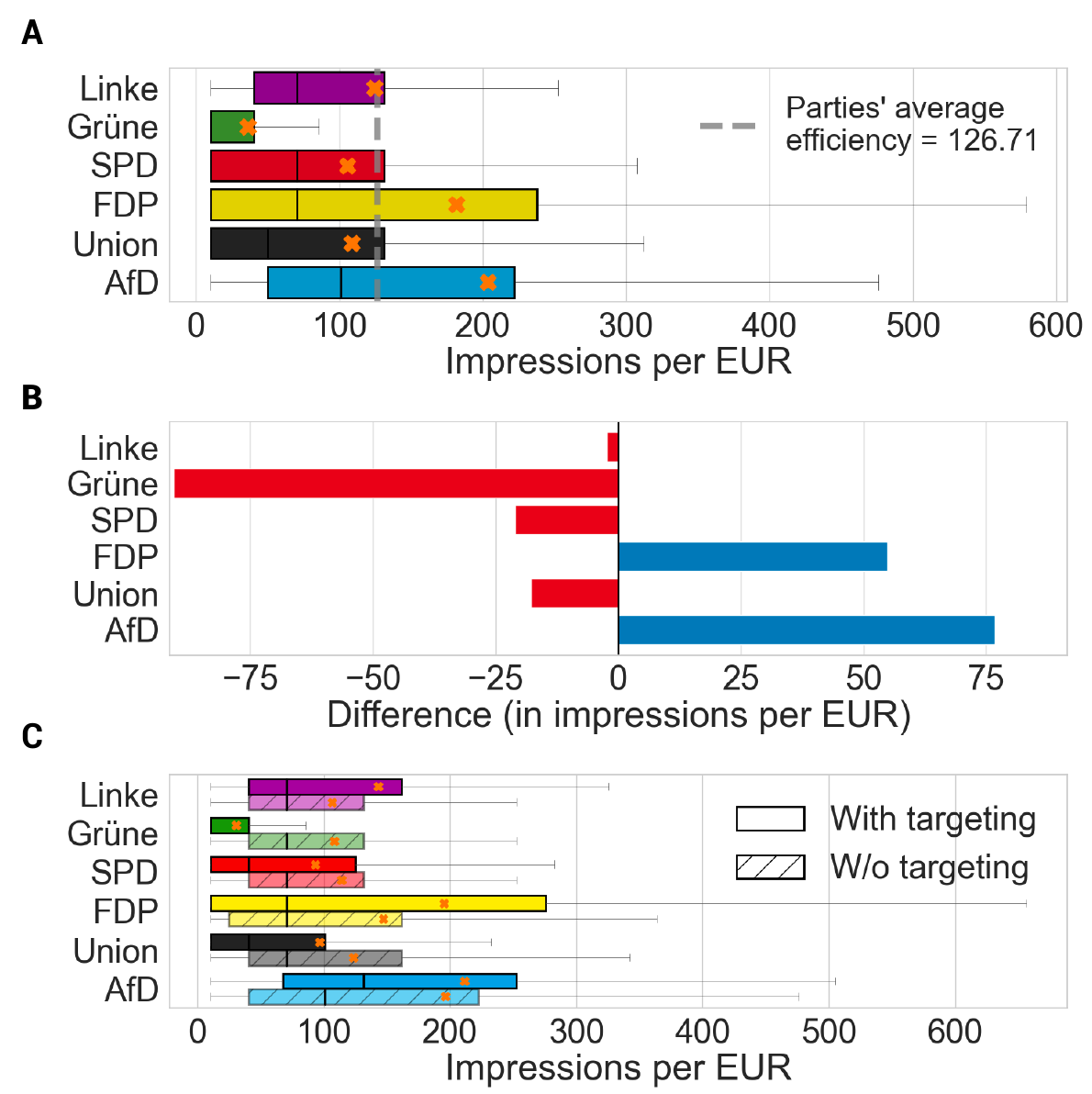}
    \caption{\figletter{a};~Distributions of impressions-per-EUR across the ads of each party. An orange cross indicates the mean of the distribution. \figletter{b};~Difference between average impressions-per-EUR of a party and average impressions-per-EUR in the overall sample. \figletter{c};~Distribution of impressions-per-EUR across the ads of each party. An orange cross indicates the mean of the distribution.}
    \label{fig:efficiency_party_difference}
\end{figure}

We now study the reach of political ads at a given budget. To do so, we focus on the number of impressions an ads generates per EUR spent (henceforth simply impressions-per-EUR). Political parties employ different campaign strategies on social media~\cite{Kruschinski.2022}, which may also lead to differences in the impressions-per-EUR of political ads. We thus compare impressions-per-EUR across political parties and test for statistically significant differences using a Kruskal-Wallis test~\cite{Kruskal.1952}. The results are in \Cref{fig:efficiency_party_difference}\figletter{a} and \figletter{b}. We find that impressions-per-EUR significantly differs across parties ($p<0.05$). On average, political ads generate \num{126.71} impressions-per-EUR. However, the \greenparty receives, on average, only \num{36.18} impressions-per-EUR.
In contrast, ads published by the \emph{FDP} and the \emph{AfD} are considerably more efficient, reaching, on average, \num{181.53} and \num{203.49} impressions per EUR, respectively. In summary, impressions-per-EUR of social media ads varies greatly across the political spectrum.

Given the high prevalence of targeted ads, we also compare impressions-per-EUR for targeted ads and ads without targeting. Our results are mixed (see \Cref{fig:efficiency_party_difference}\figletter{c}). While targeted ads achieve, on average, more impressions-per-EUR compared to ads without targeting for the \emph{Linke}, \emph{FDP}, and \emph{AfD}, the opposite is true for the \greenparty, \emph{SPD}, and \emph{Union}. Performing multiple pairwise Kruskal-Wallis tests with Benjamini-Hochberg correction based on a family-wise error rate (FWER) of $\mbox{FWER} = 0.05$, we find that these differences are all statistically significant ($p<0.05$). Generally, targeting may result in less impressions-per-EUR than ads without targeting in terms of impressions per EUR. This may be due to more narrow and thus more expensive targeting, or to an audience for which targeting is more expensive.

\newpage
\subsection*{Discrepancies between targeted and actual audiences}

Previous research has shown that algorithmic ad delivery can lead to discrepancies between intended and actual audiences on Meta \cite{Ali.2019, Ali.2021}. This may be concerning if algorithmic ad delivery would propagate existing biases in voting patterns that restrict the reach of a party among a certain population, and thus hamper fair electoral competition. For example, some parties (e.g., \greenparty) tend to receive more votes from younger female audiences while others are more popular among older male audiences (e.g., \emph{AfD}). We thus explore potential discrepancies between the demographic distribution of the intended ($=$targeted) and the actual audience reached by political campaigns due to the algorithmic ad delivery. Meta does not provide detailed information on the actual audience with respect to other targeting criteria, which is why we focus on age and gender. These demographic characteristics are nevertheless important determinants for a plethora of life outcomes and ideological stances \cite{Harting.2023}.

\begin{figure}[H]
\centering
\includegraphics[width=0.8\linewidth]{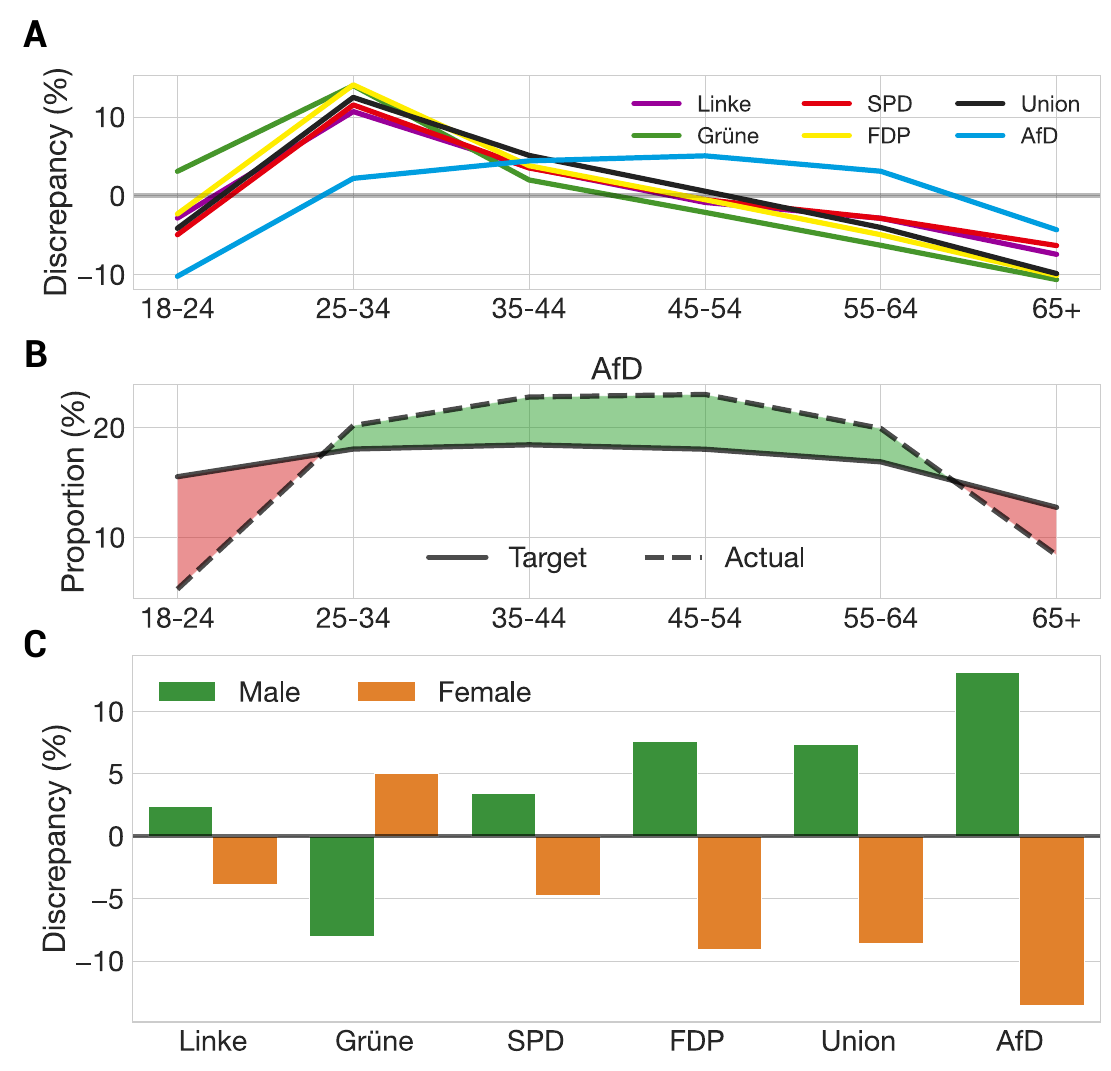}
\caption{
    \figletter{a},~Discrepancy in the age distribution between the \emph{actual} and \emph{target} audience  (in \%). We find that ads by most parties (except \emph{AfD}) are seen by more users between 25--34 than originally intended. \figletter{b},~Comparison of \emph{actual} and \emph{target} audience by age for political ads published by \emph{AfD}. A red color indicates areas where the difference between the actual and targeted audience is negative (green for positive). Younger and older users see the ads less often than originally intended by the party. \figletter{c},~Discrepancy in the gender distribution between the \emph{actual} and \emph{target} audience (in \,\%). We find large differences between male and female audiences for right-wing parties (e.g., \emph{Union}, \emph{AfD}), implying that ads are seen by considerably fewer females than originally intended due to the algorithmic ad delivery.}
\label{fig:target_actual_discrepency}
\end{figure}

Let us first focus on age (\Cref{fig:target_actual_discrepency}). For each party, we compute the proportion of actual impressions generated by different age buckets, where we weight each ad by the amount of money spent. We refer to this vector as the \emph{actual} audience of an ad. Similarly, we compute the proportion of targeted impressions by different age buckets, where we again weight each ad by the amount of money spent. This vector is the \emph{target} audience of an ad. \Cref{fig:target_actual_discrepency}\figletter{a} shows the \emph{discrepancy} (percentage difference) between the \emph{actual} and \emph{target} audience. All parties reach an actual audience that is younger than the one targeted, except for the far-right \emph{AfD} which generally reaches an older audience. We also compute the Wasserstein distance $\mathit{WS}$ (see SI for details) to quantify the difference (in years) between the age distribution of \emph{actual} and \emph{target} audiences for all parties. It yields: $\emph{Linke} = 4.69$, $\greenparty = 6.35$, $\emph{SPD} = 4.61$, $\emph{FDP} = 5.18$, $\emph{Union} = 5.69$, and $\emph{AfD} = 4.89$. Overall, the \emph{Linke} and the \emph{SPD} show the smallest discrepancy between the targeted and actual audience, while the \greenparty exhibits the largest one. \Cref{fig:target_actual_discrepency}\figletter{b} shows the values for \emph{actual} and \emph{target} for the \emph{AfD}, thereby highlighting in red (green) the age buckets where the actual audience is smaller (larger) than the targeted one.

We repeat the procedure above to compute the variables for the \emph{actual} and \emph{target} audience of an ad by gender. For all parties, there is a large \emph{discrepancy} between the targeted vs. actual audience in terms of gender, with ads shown to fewer female users than intended, except for the \greenparty (see \Cref{fig:target_actual_discrepency}\figletter{c}). This is particularly pronounced for right-wing parties such as the \emph{Union} and the \emph{AfD}. For example, the \emph{AfD} reaches 12.91\,\% more male individuals than originally targeted. Interestingly, the \greenparty is the only party for which the opposite is true: its ads reach 5\,\% more female individuals and 8\,\% fewer male individuals than targeted. To quantify the discrepancies in the gender distribution between \emph{actual} and \emph{target} audience, we again compute the Wasserstein distance for each party: $\emph{Linke} = 2.35$, $\greenparty = 6.55$, $\emph{SPD} = 3.27$, $\emph{FDP} = 7.81$, $\emph{Union} = 7.95$, and $\emph{AfD} = 13.35$, which corroborate the previous observations.

\subsection*{Systematic differences in impressions-per-EUR across parties}
\label{sec:efficiency}

\subsubsection*{\textbf{Regression analysis}}

To evaluate how key aspects of online political advertising are associated with impressions-per-EUR, we perform a regression analysis. For this, we estimate three separate linear regression models that focus on different determinants for impressions-per-EUR, namely, (1)~targeting strategies, (2)~demographics, and (3)~ad characteristics (see \nameref{sec:methods} for methodological details).

\begin{figure}
    \centering
\includegraphics[width=.65\linewidth]{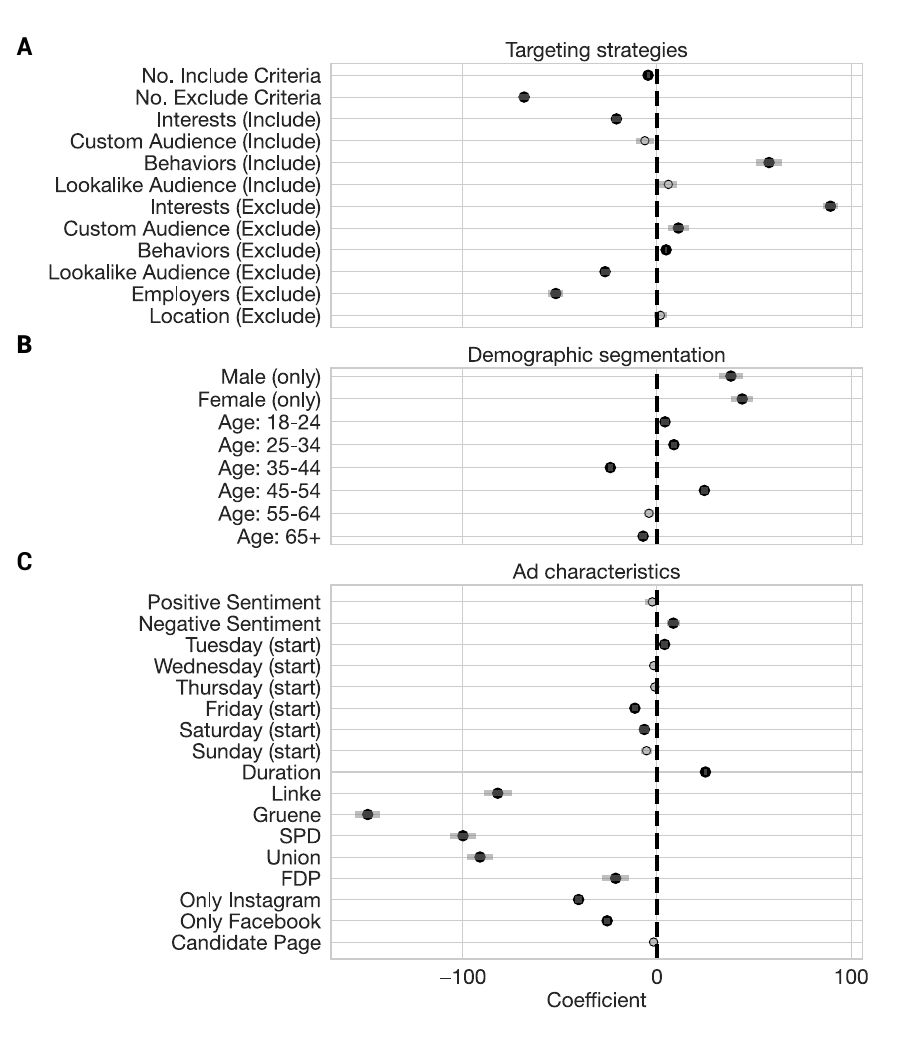}
    \caption{Coefficient estimates and 95\,\% confidence intervals for \figletter{a}~targeting strategies, \figletter{b}~demographics, and \figletter{c}~ad characteristics. Statistically significant coefficients ($p<0.05$) are indicated by black circles \tikzcircle[black, fill=black]{1.5pt} , all others by gray circles \tikzcircle[black, fill=lightgray]{1.5pt}. Sentiment, weekday, platform dummy, and party dummy are categorical variables, and the reference categories are ``neutral'' sentiment, ``Monday'', ``both platforms'', and ``AfD''.}
    \label{fig:ols-results}
\end{figure}

\emph{(1)~Targeting strategy:} Our first regression model assesses how targeting strategies are related to impressions-per-EUR (see \Cref{fig:ols-results}\figletter{a}). Ads that use more targeting criteria are linked to lower impressions-per-EUR as indicated by the negative and statistically significant coefficients for ``No. Include criteria'' ($p<0.05$) and ``No. Exclude criteria'' ($p<0.05$). This link is particularly strong for exclusions, where, all else equal, an additional 18.49 excluded criteria predicts a decrease of 68.37 impressions-per-EUR. Our regression results further show that targeting has a heterogeneous effect on impressions-per-EUR. For example, the usage of the targeting categories ``Behaviors (Include)'' and ``Interests (Exclude)'' is associated with considerably more impressions-per-EUR as indicated by a positive and significant coefficient ($p<0.05$). In contrast, the negative and significant coefficient for ``Employers (exclude)'' and ``Interests (include)'' suggest that exclusion criteria for employers or inclusion criteria for interests correspond to lower levels of impressions-per-EUR. Interestingly, excluding custom audiences is linked to higher ad efficiency, while excluding lookalike audiences is negatively associated with ad efficiency as shown by a positive and statistically significant coefficient for ``Custom audience (exclude)''  ($p<0.05$) and a negative and statistically significant for ``Lookalike (exclude)'' ($p<0.05$). Given that ad delivery heavily relies on Meta's algorithm, more transparency would be crucial to explain these findings.

\emph{(2)~Demographic segments:}
Our second regression model evaluates how different demographics explain impressions-per-EUR. The regression results are in \Cref{fig:ols-results}\figletter{b}. We find a positive and statistically significant coefficient for ``Female'' ($p<0.05$) and ``Male'' ($p<0.05$), suggesting that targeting only female or male audiences rather than all genders is associated with more impressions-per-EUR. All else equal, only targeting users from a single gender predicts, on average, an additional 43.90 (``Female only'') and 38.08 (``Male only'') impressions-per-EUR.
Furthermore, the coefficients for \emph{Age: 18--24}, \emph{Age: 25--34}, and \emph{Age: 45--54} are positive and statistically significant ($p<0.05$), while the coefficients for \emph{Age: 35--44}, and \emph{Age: 65+} are negative and statistically significant ($p<0.05$). As such, addressing younger audiences is linked to more impressions-per-EUR except for the age group between 35--44.

\emph{(3)~Ad characteristics:} In our third regression model, we study how ad characteristics are linked to impressions-per-EUR (\Cref{fig:ols-results}\figletter{c}. For example, the timing and content of an ad explain its impressions-per-EUR. In particular, ads that are online for a longer period and published earlier in the week tend to receive more impressions-per-EUR as indicated by the positive and statistically significant coefficient for \emph{Duration} ($p<0.05$) and \emph{Tuesday} ($p<0.05$) as well as the negative coefficients for the other weekdays. As such, all else equal, for each extra week that an ad remains online our model predicts 24.98 additional impressions-per-EUR. When assessing the link between impressions-per-EUR and whether a candidate has published an ad, we do not find a statistically significant coefficient ($p=0.31$). However, the party dummy is an important determinant of impressions-per-EUR. In particular, ads published by the \emph{AfD} are linked to more impressions-per-EUR compared to all other parties, which is seen by the negative and statistically significant coefficient for all other party variables ($p<0.01$). All else equal, ads by the \emph{AfD} reach +148.88, +99.86, +91.09, +81.97, and +21.25 additional impressions-per-EUR compared to the \greenparty, \emph{SPD}, \emph{Union}, \emph{Linke}, and \emph{FDP}, respectively.

Of note, sentiment, weekday, platform dummy, and party dummy are categorical variables. Hence, they need a reference condition to include them in our regression model. The coefficients of these variables should be interpreted relative to the reference categories. In this analysis, we chose ``neutral'' sentiment, ``Monday'', ``both platforms'', and ``AfD'' as reference categories.

We further conducted two additional analyses, which are motivated by the fact that larger competition may influence ad impressions. Generally, we expect competition among political ads to be small given that political ads represent only a small fraction out of all ads on Meta. Nevertheless, we perform two analyses where we control for the time-to-election-day and the number of competing political ads. We find that publishing ads earlier in the campaign is related to higher levels of impressions-per-EUR. Furthermore, higher competition in terms of more active ads at the publishing day is negatively related to impressions-per-EUR. Of note, all other results remain consistent with our main analysis except for the coefficient of negative sentiment, which is no longer statistically significant. Details are in Supplementary Material 6.

\subsubsection*{\textbf{Machine learning approach}}

We now employ a machine learning approach to predict the reach of political ads for a given budget based on the information provided by Meta. The rationale for this design is two-fold. (1)~If the available information in the Meta Ad Library and the Meta Ad Targeting Dataset is sufficiently complete, we should be able to make accurate predictions of impressions-per-EUR given the available features. In other words, a low prediction performance suggests that other unobserved features explain the heterogeneity in the reach of different ads for a given budget (e.g., factors related to the algorithmic delivery of ads), but these features are not captured in the dataset and are thus not available for external stakeholders. Hence, this can provide a glimpse into the transparency provided by the available data. (2)~The machine learning predictions can also be used to examine empirically whether there are systematic differences between the predicted and actual reach of an ad for a given budget, across different political parties. In other words, if a party consistently receives more views for a given budget than others targeting the same audience, this could indicate that the algorithmic delivery of ads is advantaging said party, which would undermine fair competition.

We fit a random forest model~\cite{Breiman.1984} by using all variables from our regression analysis related to key determinants of the reach of an ad for a given budget, namely, (1)~targeting strategies (e.g., the type and frequency of targeting categories), (2)~demographics (age, gender of target group), and (3)~ad characteristics (e.g., sentiment of an ad, ad duration, publishing party). In addition, we use the full set of targeting variables and add variables that indicate whether (i)~the advertisers supplied a data file to include/exclude a custom audience and (ii)~the data supplied by the advertisers to include/exclude a custom/lookalike audience was complete. For a full list of variables see \nameref{sec:methods}. To measure the reach of an ad for a given budget, we again rely on the number of impressions generated per EUR spent (i.e., impressions-per-EUR). Details on the implementation are in \nameref{sec:methods}.

Our model achieves an average root mean squared error $\mathit{RMSE}=123.79$ over 10 runs ($\pm$ a s.d. of 4.21) when predicting impressions-per-EUR on the hold-out set. More importantly, our model can only explain a fraction of the variance in our data ($R^2=0.40$ $\pm$ a s.d. of 0.02 over 10 runs). The low $R^2$ suggests that the available information about targeting strategies, demographics, and ad characteristics is not sufficient to fully characterize the impressions-per-EUR of the ad campaign.

\begin{figure}[H]
    \centering
\includegraphics[width=.65\linewidth]{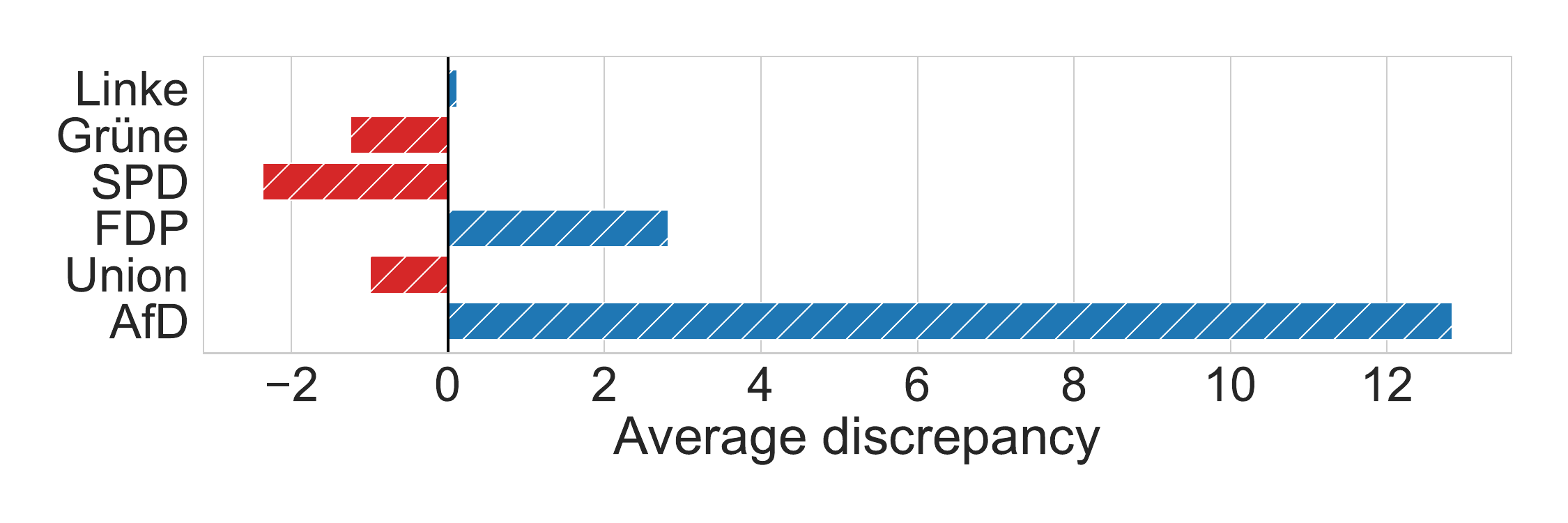}
    \caption{Average difference between actual vs. predicted impressions-per-EUR based on our machine learning model over 10 runs.}
    \label{fig:ml_results}
\end{figure}

Next, we compute the mean difference between the predicted and actual impressions-per-EUR across parties. This measure indicates whether specific parties consistently achieve more or fewer impressions-per-EUR while controlling for all other available sources of heterogeneity in targeting strategies, demographics, and ad characteristics. \Cref{fig:ml_results} shows that most left-leaning parties (i.e., the \greenparty and the \emph{SPD}, except for \emph{Linke}) and \emph{Union} consistently achieve fewer impressions-per-EUR than predicted by our model. In contrast, the \emph{FDP} and \emph{AfD}, on average, receive 12.83 and 2.81 additional impressions per EUR, respectively. This result implies a relative advantage of 10.13\,\% and 2.22\,\% in impressions-per-EUR compared to the average ad that achieves \num{126.71} impressions-per-EUR. Overall, our results suggest that the algorithmic delivery of political ads may advantage specific parties.

We perform a series of checks to ensure the robustness of our results. First, we train an XGBoost model using the same training procedure as outlined for the random forest model. Second, we check the heterogeneity of our results across platforms and re-train our random forest model by using political ads that were solely published on (i)~Facebook and (ii)~Instagram. Across both checks, our main findings regarding (1) and (2) remain consistent. Details are in Supplementary Material 10.

\section*{Discussion}

Targeted political advertising on social media has raised significant concerns regarding fairness, accountability, and transparency among researchers, policymakers, and society at large. Given the known and significant impact of political advertising on voter turnout and vote choice~\cite{Hager.2019, Coppock.2022, Aggarwal.2023, Bar.2024}, it is crucial to analyze how parties across the political spectrum employ targeting during election campaigns, and inform policymakers on the implications of targeted political advertising for democracies. 

Our findings contribute to the existing literature on political advertising on social media by providing evidence of the prevalence of targeting across the political spectrum. While social media was found to be an essential communication channel for parties during election campaigns~\cite{Fowler.2021, Capozzi.2023, Kruschinski.2022}, it was previously unclear whether and how parties make use of targeting.

Targeting was used in 72.3\,\% of all ads published on Meta during the 2021 German federal election. In fact, parties rely on a wider range of targeting categories available to advertisers on Meta. For example, a majority of campaign budget is spent for ads that exclude users based on so-called interests (e.g., social equality, environmentalism, and international relations), behaviors (e.g., early adopters of new technology, commuters, international travelers), or employers (e.g., business owners, police officers, Ford Deutschland). We further find significant differences in the reach of ads at a given budget. For example, the far-right \emph{AfD} achieves significantly more impressions-per-EUR compared to other parties. 

Our results further show that the algorithms of the platforms drive---to a large extent---\emph{who} views an ad, which can lead to discrepancies between the intended (targeted) and actual audience. Indeed, we find considerable discrepancies between both. For instance, the \greenparty reaches considerably more female users than intended, while the \emph{Union} and the \emph{AfD} reach a larger male audience than intended. Algorithmic bias in the delivery of political ads may amplify stereotypes associated with vote choices among distinct segments of the electorate. In fact, previous research has shown that algorithmic bias is responsible for systematically delivering ads based on gender stereotypes, but in a job context~\cite{Lambrecht.2019}. This aligns with the fact that the \greenparty traditionally receives more support from female voters compared to both the \emph{Union} and the \emph{AfD}, which more strongly depend on male voters.

The results of our regression analysis provides evidence for significant heterogeneity in the impressions-per-EUR of targeted ads. For example, ads targeting only a single gender achieve significantly more impressions-per-EUR. This may be due to a higher level of personalization but also raises concerns on whether algorithmic bias propagates gender-specific ads particularly well. Moreover, ads by \greenparty receive considerably fewer impressions-per-EUR than other parties. Given that their voters have an above-average income targeting them may be more costly than the average. In contrast, ads published by the \emph{AfD} generate the most impressions-per-EUR. This could be explained by the fact that incendiary political issues promoted by populist parties (e.g., anti-immigration), tend to attract high attention on social media~\cite{Bene.2022, Capozzi.2023, Klinger.2023}.

Finally, the results of our machine learning approach show that detailed information on targeting strategies, demographics, and ad characteristics cannot fully explain the reach of an ad for a given budget in terms of impressions-per-EUR. This finding indicates that current transparency measures do not suffice to evaluate how proprietary algorithms deliver political ads. In fact, we find systematic differences between predicted and actual impressions-per-EUR, which is particularly wide for the far-right \emph{AfD}. This gap is concerning as it possibly indicates the presence of algorithmic biases that may favor populist ads.

As with other research, ours is not free from limitations that offer opportunities for future work. We focus on the 2021 German federal election; extending our results to other countries is important for generalizability. Nevertheless, the federal structure in Germany is similar to other countries, and the election is comparatively large, with more than 60~million eligible voters and candidates from parties across the political spectrum~\cite{Bundeswahlleiter.2022}. Our analysis is based on a unique, large-scale dataset that is proprietary and, as such, offers granular insights. However, more research is needed to assess the accuracy of the Meta Ad Library~\cite{Edelson.2020}. Furthermore, the effects of algorithmic ad delivery are not yet fully understood and may require further data disclosure by platforms. For instance, Meta only offers specific audience details related to age and gender, without providing comprehensive pricing information. This lack of data precludes a broader analysis of how pricing mechanisms vary and the discrepancies between the targeted and actual audiences. We thus encourage future research to focus on pricing mechanisms for political ads on social media and more granular analyses of actual and targeted audiences. Lastly, while our current focus is on targeting, we also provide insights on the purpose (see Supplementary Material 4) and focus of ads (i.e., candidates vs. parties; see Supplementary Material 5). Here, interdisciplinary approaches combining computational methods and theory from political science \cite{Bar.2023} could expand our analysis on the communication strategies employed by various parties on social media beyond targeting.

Our results contribute to the political science literature by shedding light on the prevalence of targeting during real-world elections. Targeting is highly prevalent in political campaigns and allows parties to focus on specific voter segments. This result mandates additional efforts by researchers to unravel the impact of targeted political advertising on society.

Our results have further implications for parties and policymakers. For parties, we provide valuable insights into the delivery mechanisms of targeted ads on social media. For policymakers, our research emphasizes the necessity to intensify auditing and regulation regarding political advertising on social media. The discrepancies between actual and targeted audiences that we identified potentially originate from algorithmic bias that favors certain voter segments. This is concerning as such bias may harm political participation and reinforce discrimination against disadvantaged groups. 

Policymakers should pay particular attention to addressing these issues in regulatory frameworks and hold platforms accountable for ensuring fairness, accountability, and transparency in political advertising. For example, previous research \cite{Turkel.2022} recommends that platforms make adaptations to the auction mechanisms to subsidize political advertisers or provide quotas in combination with separate auctions for political and commercial advertisers to lower competition for political advertisers. Policymakers could also require randomization of ad delivery among the target population as discussed by the European Parliament \cite{CounciloftheEuropeanUnion.2023}, thereby preventing discrimination due to algorithmic bias. Most importantly, policymakers should incentivize additional transparency measures. For example, current measures are insufficient to evaluate how the platform's pricing mechanism influences ad delivery, limiting independent assessment by researchers. Mandatory disclosure of information about click rates could help to better understand the effectiveness of social media ads. Overall, there is a pressing need for policymakers to mandate such disclosures, allowing for third-party monitoring and contributing to a more accountable system of political advertising on social media.

\section*{Materials and methods}
\label{sec:methods}

\subsection*{Data}

We collect $N=\num{81549}$ political ads from the 2021 German federal election published by candidates and parties (including their youth organizations) on Facebook and Instagram. Specifically, we first obtain all ads with a starting day in the period [2021-07-01, 2021-09-26] from the \emph{Meta Ad Library API}~\cite{Meta.2022}. We then filter ads where the sponsor or page name matches any of the six major parties (i.e., \emph{Linke}, \greenparty, \emph{SPD}, \emph{FDP}, \emph{Union}, and \emph{AfD}) or one of their candidates running for office.\footnote{For detailed election results, see \url{www.bundeswahlleiterin.de/en/bundestagswahlen/2021/ergebnisse.html}} Of note, advertisers on Meta may purchase ads with the same creatives (i.e., content, image, etc.) multiple times. Therefore, we use the term `ad' to refer to a single ad contract, including its timing, budget, and targeting settings besides its content.

The Meta Ad Library API provides detailed information about content, page/sponsor name, money spent, start and stop dates, and the number of impressions distributed across gender and age (i.e., in buckets corresponding to 18--24, 25--34, 35--44, 45--54, 55--64, 65+ years). It also indicates whether an ad was published (a)~only on Facebook, (b)~only on Instagram, or (c)~simultaneously on both platforms. For ad spending and the number of impressions, Meta only reports discretized buckets. Following previous research \cite{Capozzi.2020, Fowler.2021, Capozzi.2021, Capozzi.2023, Aisenpreis.2023, Bar.2024}, we average the maximum and minimum of each bucket to obtain conservative point estimates of the ad spending and the number of impressions per ad. A full list of variables is available in Supplementary Material 7.

We further use the \emph{Meta Ad Targeting Dataset}~\cite{Meta.2023b} to access ad targeting information. The \emph{Meta Ad Targeting Dataset} contains targeting information on all social issue, electoral, and political ads published after August 3, 2020, on Facebook or Instagram \cite{Meta.2023b}. The dataset can be accessed after approval by Meta via \url{https://developers.facebook.com/docs/ad-targeting-dataset}. We accessed the dataset in August 2023 and queried targeting information for all $N=$\num{81549} political ads retrieved from the \emph{Meta Ad Library API} based on the unique ad ids. In particular, we queried information about the users' demographics (age and gender) targeted by each ad, as well as additional targeting categories such as interests, behaviors, jobs, or locations. Throughout the paper, we consider an ad to employ ``targeting'' if it uses any category provided in the dataset in addition to demographics (i.e., gender and age) and location (since the latter must always be specified in the ad creation). We further choose to ignore the ``Include location'' category, which is defined by default for every ad (e.g., ``Germany''), but we do consider ``Exclude location'' (only present for a very small minority of ads = 3.35\%). Targeting information for each ad is specified in groups of categories (e.g., ``Interests'' such as \emph{Politics} and \emph{Environment}), and the ad will include/exclude people that match at least one category in each group. We notice that the majority of the ``include'' conditions in our data contains 1-2 groups, while all ``exclude'' conditions contain a single group. For a comprehensive overview of the targeting information provided in the dataset, see Supplementary Material 8. For an example ad with targeting information see Supplementary Fig. S4.

\subsection*{Wasserstein distance}

Following \citet{Capozzi.2023}, we compute the Wasserstein distance $\mathit{WS}$ to quantify discrepancies between the demographic distribution (i.e., age and gender) of the \emph{Target} and \emph{Actual} audience of an ad. The $\mathit{WS}$ distance, also known as the earth mover's distance (EMD) or Kantorovich-Rubinstein distance, is a measure of the dissimilarity between two probability distributions. It quantifies the minimum cost required to transform one distribution into another, where the cost is determined by the amount of ``mass'' that needs to be moved from each point in one distribution to its corresponding point in the other distribution. We compute it using the \texttt{scipy} Python library, which is based on the following formula:
$$
        l_1(u, v) = \int_{-\infty}^{+\infty} |U-V|
$$
where $u$ and $v$ are the two distributions to compare, and $U$ and $V$ the respective cumulative distribution functions.

\subsection*{Regression analysis}

We hypothesize that targeting affects the impressions-per-EUR of political ads on social media. We thus use regression analysis to study what factors drive the reach of an ad at a given budget for political advertising on social media. To measure the reach of an ad, we rely on the number of impressions an ad generated per EUR spent (or simply impressions-per-EUR). Let $y_i$ denote impressions-per-EUR for an ad $i$, and let $x_i$ refer to a vector with different characteristics belonging to that ad.
We then estimate the following linear regression model
\begin{equation}\label{eqn:regression}
    y_i = \alpha + \beta^T x_i,
\end{equation}
where $\alpha$ represents the model intercept, and $\beta$ measures the associations between the variables in $x_i$ and $y_i$. For estimation, we use ordinary least squares regression (OLS) and test whether the coefficients are significantly different from zero using two-sided $t$-tests. To facilitate the interpretability of our results, we further $z$-standardize all numerical variables in $x$. Hence, we interpret our coefficients as follows: For numerical variables, a one standard deviation increase in $x$ predicts an increase in the reach of an ad at a given budget by $\beta$ impressions-per-EUR. For dummy variables, a dummy variable in $x$ set to one predicts an increase in the reach of an ad at a given budget by $\beta$ impressions-per-EUR.

We use three different regression models to analyze how impressions-per-EUR varies across ads. To do so, we use a comprehensive set of variables available to us in the Meta Ad Library or the Meta Ad Targeting Dataset that represent three key determinants of impressions-per-EUR. In particular, we focus on (1)~different targeting strategies, (2)~different demographics, and (3)~different ad characteristics. We estimate separate models for each of these dimensions to avoid multicollinearity~\cite{Prollochs.2021} and facilitate interpretability.

\begin{squishlist}
\item[\emph{Targeting strategy:}]
To assess the role of targeting strategies as a determinant of impressions-per-EUR, we compute two variables that correspond to the overall number of inclusion and exclusion criteria. The hypothesis is that more granular targeting may incur higher costs. Further, we consider which targeting categories have been used by a party. Here, we include two variables for each targeting category available to advertisers on Meta that indicate whether a certain category was used ($=1$ if the category was used, and $=0$ otherwise) to distribute an ad. The first (second) variable indicates whether users should be included (excluded) from the audience based on the corresponding targeting category. Supplementary Table S3 lists all targeting categories available to advertisers on Meta. Targeting categories are often employed together. To improve interpretaiblity and mitigate multicollinearity concerns, we thus focus on the top-10 targeting categories (by total spending). The top-10 targeting categories (by total spending) are shown in Supplementary Figure S1.

\item[\emph{Demographics:}]
To study which demographics are linked to higher impressions-per-EUR, we include whether an ad is targeting only users from a single gender (i.e., only female or male users; $=1$ if yes, and $=0$ otherwise) as well as the share of targeted users across different age groups (i.e., 18--24, 25--34, 35--44, 45--54, 55--64, 65+).

\item[\emph{Ad characteristics:}]
Ad characteristics are likely to influence the reach of political ads at a given budget. Hence, we study how ($i$)~content, ($ii$)~timing, ($iii$)~the platform an ad was published on, and ($iv$)~the publisher of an ad is linked to impressions-per-EUR. ($i$)~\emph{Content}: We thus analyze the sentiment conveyed by an ad and classify the content of each ad as ``positive'', ``neutral'', or ``negative''. We use German Sentiment Bert, a state-of-the-art transformer-based sentiment model for German text that was trained on 5.4 million labeled samples~\cite{Guhr.2020}. ($ii$)~\emph{Timing}: We analyze whether launching an ad on specific weekdays is beneficial and how ad duration (i.e., the timespan an ad remains online) relates to impressions-per-EUR. ($iii$)~\emph{Platform:} Platform characteristics such as audience, user behavior, and ad competition can influence political campaigning~\cite{Bossetta.2018}. Hence, we study the association between impressions-per-EUR and the platform on which an ad was published, by using dummy variables to encode whether an ad was published on Facebook, Instagram, or both platforms simultaneously. ($iv$)~\emph{Publisher}: In the context of the German dual-vote system, which features both party and candidate votes, political science literature shows the role of both party and candidate behavior in shaping voter perceptions~\cite{Gschwend.2015}. These perceptions may affect the impressions-per-EUR of ads authored by different party and candidate pages. Thus, we encode the different parties using dummy variables. We further encode whether the ad was distributed through a candidate's page ($=1$ if yes, and $=0$ otherwise).

Sentiment, weekday, platform dummy, and party dummy are multi-level categorical variables. Hence, we have to choose a reference condition to include them in our regression model and interpret the coefficients relative to the reference categories. In analysis, we choose ``neutral'' sentiment, ``Monday'', ``both platforms'', and ``AfD'' as reference categories.
\end{squishlist}

\subsection*{Machine learning approach}

\textbf{Objective:} We employ a machine learning approach to evaluate whether we can predict the the reach of an ad for a given budget as measured by impressions-per-EUR. The aim is two-fold. (1)~We study whether a machine learning model can accurately predict impressions-per-EUR of ads based on the information provided in the Meta Ad Library and Meta Ad Targeting Dataset. A high prediction performance implies that the provided information is sufficient to understand the ad delivery mechanisms of the platforms. Vice versa, a low prediction performance suggests the presence of unobserved confounders that explain the observed heterogeneity but which are currently unavailable for external audits. (2)~We analyze the difference between predicted and actual impressions-per-EUR across parties. In a fair environment, we would expect no systematic differences in terms of impressions-per-EUR between different parties. In contrast, if a party consistently receives more views at a given budget than others targeting the same audience, this could indicate that the algorithmic delivery of ads is advantaging said party, which would undermine fair competition.

\vspace{0.2cm}
\noindent\textbf{Features:} For our machine learning model, we use all variables from our regression model that represent key determinants of impressions-per-EUR: (1)~different targeting strategies, (2)~different demographics, and (3)~different ad characteristics (see above).

In addition, we use the full set of targeting variables (see Supplementary Table S3 for a full list) and add variables that indicate whether (i)~the advertisers supplied a data file to include/exclude a custom audience ($=1$ if the a data file was supplied, and $=0$ otherwise) and (ii)~the data supplied by the advertisers to include/exclude a custom/lookalike audience was complete ($=1$ if the data was complete, and = 0 otherwise).

\noindent\textbf{Implementation:} We (1)~study whether machine learning can accurately predict impressions-per-EUR based on the above variables and (2)~analyze the difference between predicted and actual impressions-per-EUR. To do so, we fit a random forest model~\cite{Breiman.1984} by using all features from above. We split our data into a training (80\,\%) and a hold-out set (20\,\%) for evaluation, and tune the model via 10-fold cross-validation in combination with a grid search (see below). To control for the number of ads published by each party, we weigh observations inverse proportionally to the ad frequency by party in our training set when fitting the model. We further $z$-standardize numeric variables.

\vspace{0.2cm}
\noindent\textbf{Hyperparameter tuning:} For the training of the random forest model, we use 10-fold cross-validation in combination with a grid search. In particular, we vary (1)~the number of trees used for a forest (\emph{$N$ estimators}), (2)~the number of variables to consider at each split (\emph{Max features}), (3)~maximum depth of the tree (\emph{Max depth}), (4)~minimum samples to split a node (\emph{Min node}), (5)~minimum samples in a leaf (\emph{Min leaf}), and (6)~whether to bootstrap samples when building trees (\emph{Bootstrap}). Details on the hyperparameter tuning are in Supplementary Material 9.

\vspace{0.2cm}
\noindent\textbf{Evaluation:} We evaluate the predictive power of our model based on the average root mean squared error (RMSE) over 10 runs with different seeds. We further rely on the $R^2$ to determine how well our set of variables is able to explain variance in impressions-per-EUR. We compute the $R^2$ via
\begin{equation}
R^2 = \frac{\sum_{i=1}^{n}(y_i - \bar{y})(x_i - \bar{x})}{\sqrt{\sum_{i=1}^{n}(y_i - \bar{y})^2}\sqrt{\sum_{i=1}^{n}(x_i - \bar{x})^2}},
\end{equation}
where $n$ is the number of observations, with $x_i$ referring to a vector with different ad characteristics and $y_i$ representing predicted impressions-per-EUR, and $\bar{y}$ and $\bar{x}$ as their respective means. In a fully transparent setting, we would expect the $R^2$ to be close to 1.

\newpage

\section*{Author contributions} 

DB, FP, GDFM, and SF contributed to conceptualization. DB and FP contributed to data analysis. DB, FP, GDFM, and SF contributed to results interpretation and manuscript writing. DB, FP, GDFM, and SF approved the manuscript.

\section*{Competing interests}

The authors declare no competing interests.

\section*{Data availability statement}

\noindent \textbf{Code availability:} All code to replicate our analyses is available via our GitHub repository at \url{https://github.com/DominikBaer95/auditing_targeted_political_advertising}.

\noindent \textbf{Data availability:} All data used for the analysis is publicly available. Data on political ads on Facebook and Instagram is available via the Meta Ad Library: \url{https://www.facebook.com/ads/library/}. Targeting data is available via the Meta Ad Targeting Dataset: \url{https://developers.facebook.com/docs/fort-ads-targeting-dataset/}. To ensure reproducibility, we provide ids for all ads in our dataset, which can be used to retrieve the original data through Meta Ad Library (both through the API and with the web interface by simply using the id as query) via our GitHub repository at \url{https://github.com/DominikBaer95/auditing_targeted_political_advertising}. Due to Meta's ToS we cannot share any further information.

\section*{Funding information}
FP is partially funded by the European Union (NextGenerationEU project PNRR-PE-AI FAIR), the Italian Ministry of Education (PRIN PNRR grant CODE and PRIN grant DEMON). This manuscript reflects only the authors’ views and opinions, and funding bodies are not responsible for them.

\newpage
\bibliographystyle{ACM-Reference-Format}
\bibliography{literature.bib}

\newpage 
\appendix

%%%%%%%%%%%%%%%%%%%%%%%%%%%%%%%%%%%%%%%%%%%%%%%%%%%%%%%%%%%%%%%%%%%%%%%%%%%%%%
% Title page

\begin{center}
\vspace{3cm}
{\huge\singlespacing Systematic discrepancies in the delivery of political ads on Facebook and Instagram}

\vspace{0.8cm}

\huge Supplementary Materials

\end{center}

\thispagestyle{empty}

%%%%%%%%%%%%%%%%%%%%%%%%%%%%%%%%%%%%%%%%%%%%%%%%%%%%%%%%%%%%%%%%%%%%%%%%%%%%%%

\renewcommand{\thesection}{\arabic{section}}
\renewcommand{\tablename}{Table}
\renewcommand{\thetable}{S\arabic{table}}
\renewcommand{\figurename}{Figure}
\renewcommand{\thefigure}{S\arabic{figure}}
\setcounter{figure}{0}
\setcounter{table}{0}

\newpage
\section{Related work}
\label{supp:related_work}

% Political advertising and social media
Social media has led to a major shift in political advertising~\cite{Fowler.2020, Fowler.2021, Sosnovik.2021}. Due to the large user base of social media platforms, advertisers are able to run campaigns with wide reach at comparatively low costs~\cite{Fowler.2021}. This capability helps campaigns with smaller budgets and may democratize elections by fostering electoral competition~\cite{Fowler.2020, Fowler.2021}. However, the use of social media for political advertising has also introduced new challenges. For example, political ads on social media tend to be more partisan compared to traditional forms like television ads~\cite{Fowler.2021}. Additionally, far-right and populist parties, known for promoting anti-democratic narratives, appear to benefit from advertising on social media~\cite{Capozzi.2023} and even use it to spread misinformation~\cite{CanoOron.2021}. Given that political advertising was shown to influence voter turnout and vote choice~\cite{Hager.2019, Coppock.2022, Aggarwal.2023}, it is crucial to audit political advertising on social media in order to ensure the fairness, accountability, and transparency of electoral processes.

%Political parties publish content on Facebook to gain visibility among the electorate and to engage their voters. This content includes both advertising and organic content, with lines between these categories blurred by `boosted' posts. An important feature for the success of political posts on Facebook is interactivity: responsive party posts garner more shares, likes, and comments \cite{KocMichalska.2021}. For parties, these results provide useful insights into what makes for an efficient Facebook campaign in terms of how they can accelerate the reach of their communication. Advertising, however, does not allow for this type of interaction. Instead, the audience reach has been studied for this type of content, finding that the political parties advertising on an anti-immigration platform in Italy reach voters similar in age and gender to their voter base \cite{Capozzi.2020, Capozzi.2023}. We find a similar result for Germany in our work, this time by looking at the targeting logic used by the parties themselves.

% Targeted advertising
An important benefit of advertising on social media is targeting.
Targeting allows advertisers to craft customized messages directed at specific user groups, which makes it particularly valuable for political parties during election campaigns~\cite{Fowler.2021}. In fact, targeted advertising has been shown to be effective in various settings outside of elections. Examples include increasing conversion rates for products~\cite{Matz.2017}, promoting public health measures to reduce COVID-19 infection rates~\cite{Breza.2021}, and shifting views on climate change~\cite{Goldberg.2021}. Furthermore, previous research has shown that parties strategically tailor social media ads to target specific demographic groups~\cite{Erfort.2023}. However, the prevalence of targeted political ads as well as the detailed targeting strategies across the political spectrum remain unclear. To fill this gap, we provide a comprehensive analysis of targeted advertising on social media 

% Criticism and why auditing is necessary
Auditing online political advertising follows a long tradition of scrutiny in our democratic process~\cite{Edelson.2019, Edelson.2020, Ali.2021}. A major concern is that hyper-personalization could lead to political filter bubbles and echo chambers~\cite{Cinelli.2021, Garimella.2018, Ali.2021}. Moreover, concerns have been raised that targeted advertising may discriminate against parts of the electorate~\cite{Speicher.2018} or use personally sensitive data (e.g., ethnic origin, sexual orientation) to identify receptive audiences~\cite{Cabanas.2021}. In fact, previous research has shown that the algorithmic delivery of ads discriminates against women \cite{Lambrecht.2019}, which may reinforce existing stereotypes and harm political participation. Overall, this underlines the need for a better understanding of political advertising on social media (as a first step to developing regulatory frameworks).

%around ``lookalike'' audience matching, which may be used for discriminatory advertising \cite{Speicher.2018}. This feature allows the advertiser to define the audience by supplying a source list of users, and thus leveraging potentially sensitive attributes (such as ethnic affinity). Facebook specifically has been blamed for promoting divisive content harmful to the democratic process \cite{Vaidhyanathan.2017, Eady.2023}. For instance, \citet{Mejova.2020} found divisive messaging in Facebook ads about the COVID-19 vaccine and other preventive measures which competed with messaging by public health institutions. Further, \citet{CanoOron.2021} found misinformation in the advertising by VOX, a far-right populist Spanish party.

%Systemically, social media platforms play an increasingly important role in the political messaging ecosystem, thus giving rise to what \citet{Woolley.2019} call ``computational propaganda''---a combination of manual curation and algorithmic optimization to maximize the propagation of a message. 

%However, previous research on the Ad Library has shown that Meta's disclosure of political advertising activity is not always accurate \cite{Edelson.2020}. 
%Indeed, instances of undeclared coordinated activity by ``inauthentic communities'' were found, which funded large-scale advertising campaigns.

% Community-driven auditing
There have been community-driven efforts to independently monitor political ads online and thus improve the transparency of political advertising~\cite{Silva.2020, Matias.2022}. These systems often rely on data donations~\cite{Silva.2020}, or volunteers to audit political advertising on social media platforms~\cite{Matias.2022}. As such, they are independent from platforms and offer a transparent view on online political advertising. However, such efforts are unable to monitor political advertising at scale, are limited to information that is publicly available on the platforms (e.g., cannot capture ad spending), and are biased toward the community and thus not representative. To address these issues, we leverage a novel dataset that offers in-depth insights into targeting strategies for political ads.

% Platform oriented auditing
Following public pressure and regulatory efforts, platforms have started to release internal data that records political ads at scale and provides comprehensive insights into advertising behavior beyond the scope of community-driven systems, including information about actual spending and real-world impressions. For example, the Meta Ad Library provides public access to all political ads published on Facebook and Instagram. Researchers have used these resources to study how politicians advertise on climate change~\cite{Aisenpreis.2023} and immigration~\cite{Capozzi.2020, Capozzi.2021}, address Spanish vs. English-speaking audiences during the 2020 U.S. election~\cite{Coelho.2023}, and analyze political ads by populist and mainstream parties during the 2019 European elections~\cite{Capozzi.2023} and the 2022 Italian election~\cite{Pierri.2023}. However, a significant gap exists in our understanding of the targeting strategies employed in political advertising. To close this gap, our study offers the first analysis of \emph{targeting} for political advertising on social media.

%For instance, \citet{Matias.2022} designed a software-supported approach for auditing, which uses coordinated volunteers to analyze political advertising policies enacted by Facebook and Google during the 2018 U.S. election. A team of volunteers analyzed the companies' reactions to posting auto-generated ads and found systematic errors in how they enforced policies.

Given the importance of algorithmic ad delivery for targeting, researchers and policymakers are worried that safeguarding the integrity of the democratic process is now in the hands of commercial actors, ``who may have differing understandings of fundamental democratic norms'' \cite{Dommett.2019}. Beyond researchers and policymakers, a large part of the public is also concerned about targeted political advertising: according to the Pew Research Center, more than half of the adult U.S. population finds that social media platforms should ban political advertising, and more than three-quarters find that targeting for political campaigns is not acceptable \cite{Auxier.2020}. Therefore, auditing is crucial to ensure fairness, accountability, and transparency in electoral processes.

\newpage
\section{The 2021 German federal election}
\label{supp:background_election}

% Overview
On September~26, 2021, more than 60 million Germans were called to elect a new parliament (called ``Bundestag'')~\cite{Bundeswahlleiter.2022}.
The election marked a turning point in German politics as the previous chancellor, Angela Merkel, did not stand for re-election after 16 years~\cite{TheEconomist.2023}.

%and three parties nominated a candidate for chancellor for the first time. As such, the election was considered particularly competitive which may have led to additional campaign efforts.

% Electoral system
In Germany, each voter casts two votes: The \emph{first vote} (``Erststimme'') selects a preferred candidate as a constituency representative, where a majority vote determines who secures a seat in the parliament. The \emph{second vote} (``Zweitstimme'') is for a political party, for which seats in the parliament are allocated based on each party's share of second votes. The dual-vote system in Germany ensures both individual representation and party influence. Furthermore, due to the dual-vote system, election campaigns in Germany have to simultaneously appeal to the local concerns of voters and broader national issues. While candidates may run their own campaigns, overarching party campaigns typically dominate~\cite{Gschwend.2015}.

% Parties
The German multi-party system has six major political parties that compete for voters across the political spectrum~\cite{DeutscherBundestag.2023}\footnote{There are further candidates running for smaller parties or as independents. However, smaller parties and independents play only a minor role given their limited resources and particularities of the German electoral system in that parties need a second vote share of at least 5\,\% or win three constituencies to gain seats in the parliament. This, in turn, means that it is intentionally made highly unlikely for smaller parties and independents to enter the parliament. Hence, we focus on political ads by the six main parties---\emph{Linke}, \greenparty, \emph{SPD}, \emph{FDP}, \emph{Union}, and \emph{AfD}---throughout our paper. These parties have also been part of the parliament in the legislative period before the 2021 election.}: (1)~\textbf{\textcolor{linkecolor}{\emph{Linke}}} (\emph{The Left}) is a democratic socialist party located on the political left that promotes progressive social and economic policies. (2)~\textbf{\textcolor{greenscolor}{\greenparty}} (\emph{Alliance 90/The Greens}) is located at the center-left of the political spectrum with a strong emphasis on environmental topics and social equality, because of which the party is particularly popular among young urban voters. (3)~\textbf{\textcolor{spdcolor}{\emph{SPD}}} (\emph{The Social Democratic Party of Germany}) is also located at the center-left of the political spectrum and traditionally focuses on social equality. (4)~\textbf{\textcolor{fdpcolor}{\emph{FDP}}} (\emph{The Free Democratic Party}) is a liberal party located in the center of the political spectrum advocating a liberal and market-oriented agenda that gained large support from young voters. (5)~\textbf{\textcolor{unioncolor}{\emph{Union}}} (The \emph{Union}) is the main center-right party in Germany. It is a coalition of the \emph{CDU} (\emph{Christian Democratic Union}), which operates in all federal states except Bavaria, and the \emph{CSU} (\emph{Christian Social Union}, which only operates in Bavaria). The \emph{Union} promotes conservative values and supports a market-oriented economy with a balanced approach to social policies and Christian values. (6)~\textbf{\textcolor{afdcolor}{\emph{AfD}}} (\emph{The Alternative for Germany}) was founded in 2013 and is a far-right party with a strong focus on immigration and public security. It is popular among an older and male-dominated electorate. The \emph{AfD} has been a source of contention in German society, with criticisms highlighting its anti-immigrant rhetoric, affiliations with far-right extremism, and tendencies towards historical revisionism. The party's divisive stances and oversimplification of complex issues from a populist perspective have led to a situation where the \emph{AfD} is politically isolated from the other main parties.

\newpage
\section{Political advertising on Meta during the 2021 German federal election}
\label{supp:descriptives_data}

Our data comprises $N=\num{81549}$ ads with an overall cost of EUR~$9.8$ million for more than $1.1$ billion impressions. Generally, larger parties ran a larger number of ads. However, we also see two exceptions: the winning party (\emph{SPD}) was not particularly active ($\sim16$k ads), while one of the smaller parties (\greenparty) ran a disproportionately high number of ads (over $39$k). In \Cref{tab:description_dataset}, we show a breakdown in terms of ads, money, and impressions for all parties.

% descriptive table
\begin{table}[H]
    \centering
    \sisetup{table-alignment=right,table-format=9.0}
    \caption{Breakdown of our dataset in terms of the total number of ads, money spent, and impressions generated by each party. Money and impressions data come in brackets; for closed ranges, we consider the average of the endpoints of the range, and, for open-ended ranges, we take the known closed endpoint.}
    \label{tab:description_dataset}
    \footnotesize
    \begin{tabular}{lSSS}
    \toprule
    Party & {Number of ads} & {Spending (EUR)} & {Impressions} \\
    \midrule
    \emph{Linke} & \num{2257} & \num{390950} & \num{75087865} \\
    \greenparty & \num{38604}& \num{3608103} & \num{ 205728143} \\
    \emph{SPD} & \num{12525} & \num{1526741} & \num{169188312} \\
    \emph{FDP} & \num{10327} & \num{1369329} & \num{319367015} \\
    \emph{Union} & \num{15349} & \num{2337002} & \num{299763160} \\
    \emph{AfD} & \num{2487} & \num{577429} & \num{87361955} \\
    \midrule
    Total & \num{81549} & \num{9809553}  & \num{1156496450} \\
    \bottomrule
    \end{tabular}
\end{table}

\begin{figure}
    \centering
    \includegraphics[width=\linewidth]{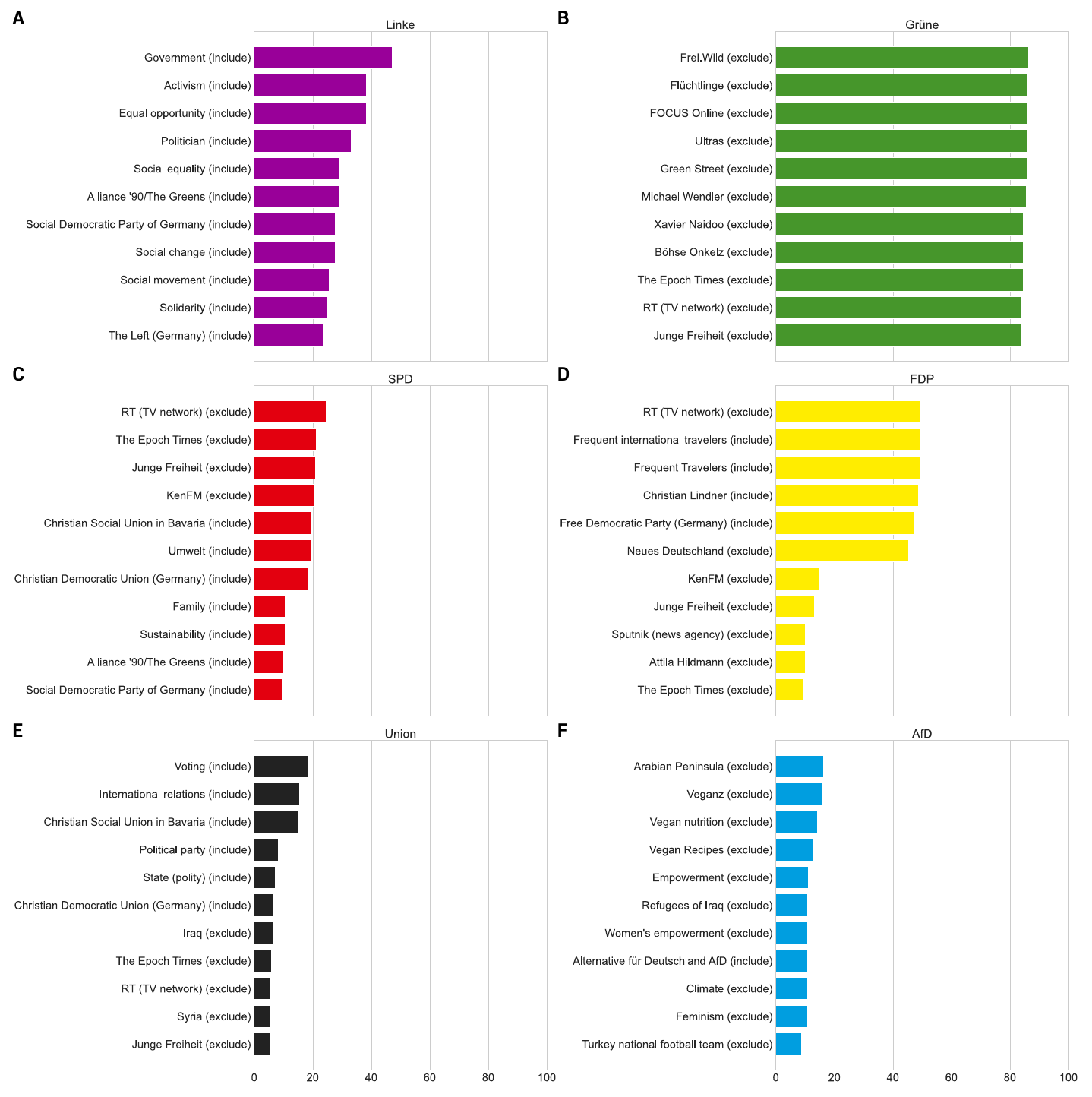}
    \caption{Top-10 targeting criteria employed by \figletter{a}~\emph{Linke}, \figletter{b}~\greenparty, \figletter{c}~\emph{SPD}, \figletter{d}~\emph{FDP}, \figletter{e}~\emph{Union}, and \figletter{f}~\emph{AfD} in terms of overall spending (in EUR).}
    \label{fig:top-criteria-parties}
\end{figure}

\Cref{fig:top-criteria-parties} shows the top-10 targeting criteria employed by different parties in terms of spending.

%\Cref{fig:targeting_categories_length} shows the distribution of the number of unique targeting categories (for Include and Exclude conditions) used by each ad, for all parties.

% \begin{figure}[!t]
%     \centering
%     \includegraphics[width=.65\linewidth]{img/targeting_categories_length.pdf}
%     \caption{Distributions of the number of different targeting categories used by parties. An orange cross indicates the mean of the distribution. %Outliers (i.e.,values outside $1.5\times$ the inter-quartile range) are omitted.
%     }
%     \label{fig:targeting_categories_length}
% \end{figure}

\clearpage
\section{Goals of political advertising on social media}
\label{supp:purpose_ads}

Political campaigning follows various purposes \cite{Ballard.2016, Zhang.2017, StromerGalley.2021, Ridout.2021}. To explore the purpose of political ads on social media, we conduct an additional qualitative analysis. Specifically, we manually analyzed and coded a random sample of $N=100$ ads for their purpose. Inspired by political science literature \cite{Ballard.2016, Zhang.2017, StromerGalley.2021, Ridout.2021}, we distinguish ads that mobilize, persuade, promote events, or call for donations. Furthermore, we distinguish whether an ad is informational (i.e., purely focusing on explaining a policy) or attacking opponents.

%We have also considered computational approaches to provide a more comprehensive analysis. However, we realized the diversity and nuanced idiosyncrasies in political advertising can only be captured in a qualitative investigation. 

Our analysis shows that a large majority of ads are designed to mobilize (50\,\%) or persuade (31\,\%) voters of a political party or candidate. Few ads are informational (9\,\%) or promote events (8\,\%). We find only 2 ads attacking another party, while no ad is calling for donations. The latter is not surprising since parties in Germany are mostly funded publicly or via membership fees and would not call for donations during electoral campaigns.

Overall, our analysis suggests that political ads on social media are designed for various purposes but predominantly focus on mobilization and persuasion of voters.

\clearpage

\section{Political advertising by parties vs. candidates}
\label{supp:party_vs_candidates}

In the German voting system, voters cast two votes: The \emph{first vote} (``Erststimme'') selects a preferred candidate as a constituency representative and the \emph{second vote} (``Zweitstimme'') is for a political party (see \Cref{supp:background_election} for details). As such, political campaigns may focus on parties more broadly but also specific candidates.

To check whether the ads in our sample are focusing on parties more broadly or tend to focus on specific candidates we ran two additional analysis. First, we classify ads by whether they have been published on pages run by candidates. This follows the rationale that ads published by candidates are likely to be tailored to the specific candidate. In contrast, parties are likely to advertise more broadly. We collected all candidate names for the 2021 German federal election from the Federal Returning Officer \cite{Bundeswahlleiter.2022} and matched the candidate names with the page name of an ad. We find that $18135$ ads (i.e., 22.24\,\% of all ads in our sample) are published by candidates and thus likely to specifically focus on a candidate. Second, since it is likely that parties sometimes also specifically advertise for a candidate, we employed a qualitative approach and manually classified a random sample of $N=100$ ads to check whether ads tend to be focused on specific candidates or parties broadly. We find that 59\,\% of ads are focused more broadly on a party while 41\,\% tend to focus on a specific candidate. We further find that all ads published on a candidate page are also focusing on the specific candidate thus corroborating our first analysis.

Overall, we find that ads published by parties tend to be focused more broadly on the party while ads published by candidates tend to focus on the specific candidates.

\clearpage
\section{The role of competition on impressions-per-EUR}
\label{supp:timing}

We conducted two additional analyses to study the role of competition. Due to the absence of other variables on competition at Meta, we focus on two variables: the timing-to-election-day and the number of competing political ads.

\begin{figure}[H]
    \centering
\includegraphics[width=.65\linewidth]{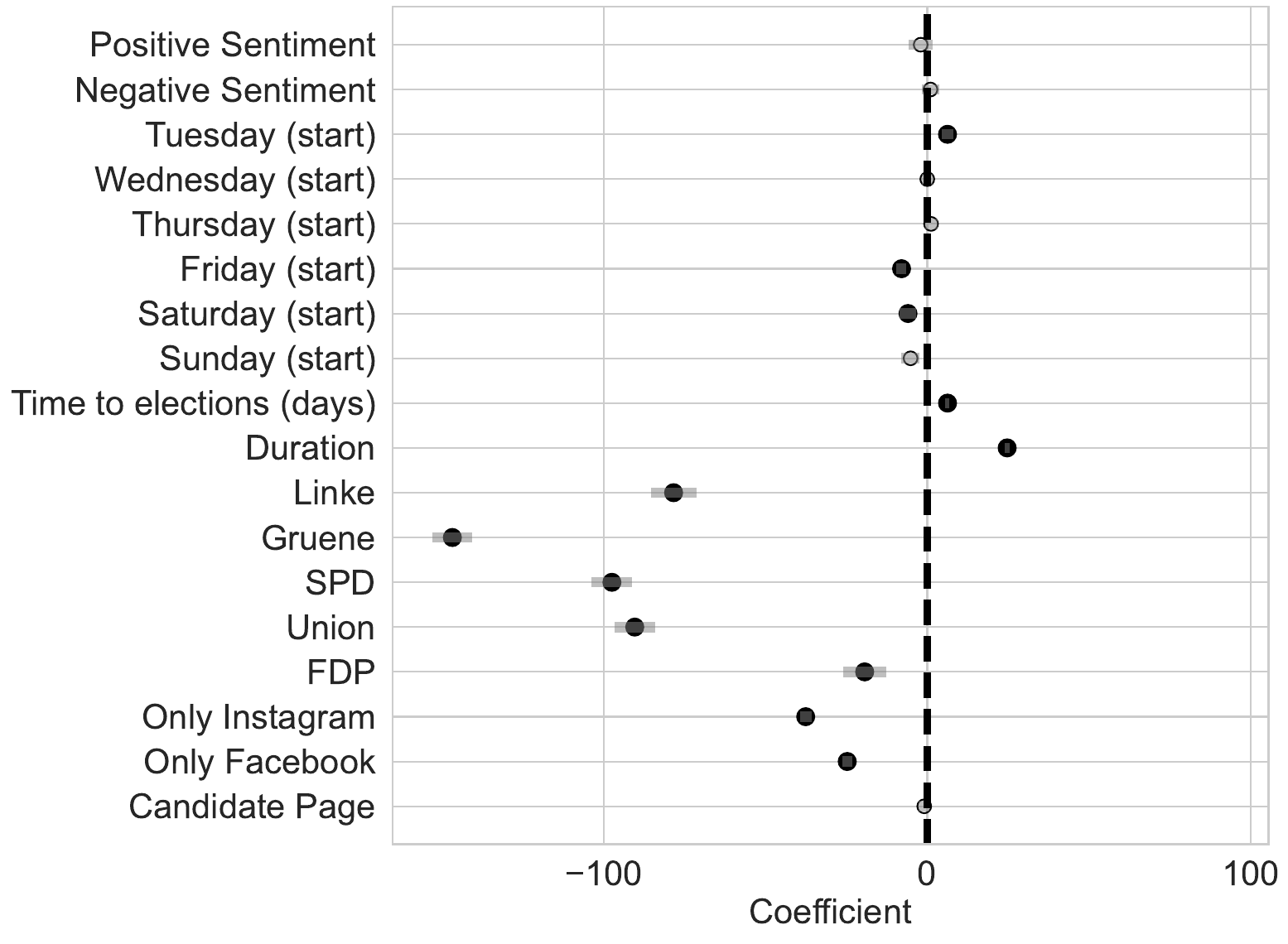}
    \caption{Coefficient estimates and 95\,\% confidence intervals for ad characteristics including the time to election in days (\emph{Time to elections (days)}). Statistically significant coefficients ($p<0.05$) are indicated by black circles \tikzcircle[black, fill=black]{1.5pt} , all others by gray circles \tikzcircle[black, fill=lightgray]{1.5pt}.Sentiment, weekday, platform dummy, and party dummy are categorical variables, and the reference categories are ``neutral'' sentiment, ``Monday'', ``both platforms'', and ``AfD''.}
    \label{fig:ols-results_time}
\end{figure}

For the first analysis, we re-run the regression model for ad characteristics from our main analysis but included an additional variable measuring the time between publication and election day. The results are in \Cref{fig:ols-results_time}. We find a positive and statistically significant coefficient for \emph{Time to election (days)} ($p<0.05$), suggesting that publishing ads earlier in the campaign is related to higher levels of impressions-per-EUR. All else equal, ads published one standard deviation of days earlier receive, on average, 6.26 additional impressions-per-EUR. This may be due to less competition for political audiences at the beginning of the campaign peri

\begin{figure}[H]
    \centering
\includegraphics[width=.65\linewidth]{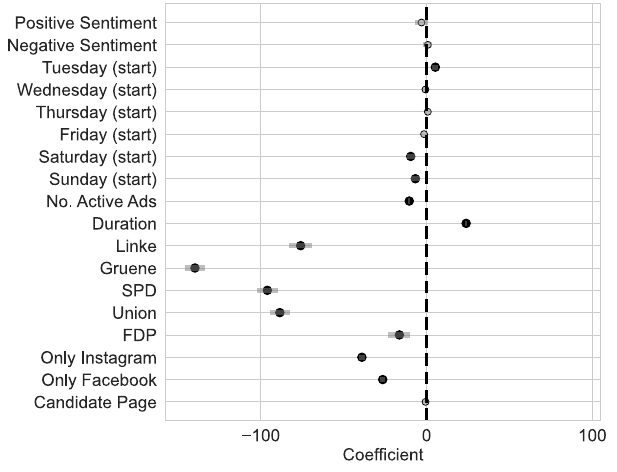}
    \caption{Coefficient estimates and 95\,\% confidence intervals for ad characteristics including the the number of active ads at the day an ad was first published (\emph{No. Active Ads}). Statistically significant coefficients ($p<0.05$) are indicated by black circles \tikzcircle[black, fill=black]{1.5pt} , all others by gray circles \tikzcircle[black, fill=lightgray]{1.5pt}. Sentiment, weekday, platform dummy, and party dummy are categorical variables, and the reference categories are ``neutral'' sentiment, ``Monday'', ``both platforms'', and ``AfD''.}
    \label{fig:ols-results_competition}
\end{figure}

Our first analysis indicates that publishing ads earlier during the campaign is linked to higher levels of impressions-per-EUR. While this may be related to lower competition, it could also be due to other idiosyncrasies pertaining to the specific time an ad was published unrelated to competition. Hence, we also studied how the total number of active political ads (across all parties) on the day an ad was published is linked to impressions-per-EUR as a more appropriate measure of competition. To do so, we again re-run the regression model for ad characteristics from our main analysis but now included an additional variable measuring the number of active ads on the publishing day of an ad. The results are in \Cref{fig:ols-results_competition}. We find a negative and statistically significant coefficient for \emph{No. Active Ads} ($p<0.05$), indicating that higher competition is related to less impressions-per-EUR. All else equal, a one-standard-deviation increase in competing political ads ($\approx4000$ ads) is related to a decrease of 10.45 impressions-per-EUR.

Of note, for both analyses, all other results remain consistent with our main analysis except for the coefficient of negative sentiment, which is no longer statistically significant.

\clearpage

\clearpage
\section{Meta ad library}
\label{supp:meta_ad_library}

The Meta Ad Library provides the variables as described in \Cref{tab:ad-library-variables} for ``ads about social issues, election or politics that were delivered anywhere in the world during the past seven years''. We refer the reader to the API documentation (\url{https://www.facebook.com/ads/library/api/}) for further details.

\begin{table}[!h]
\caption{List of variables provided in the Meta Ad Library API.}
\footnotesize
%\ttfamily
\centering
\begin{tabular}{l}
\toprule
    Creation time \\ \midrule
    Text \\ \midrule
    Link descriptions, captions and titles \\ \midrule
    Start and stop time \\ \midrule
    Snapshot URL \\ \midrule
    Sponsor name \\ \midrule
    Regional distribution \\ \midrule
    Demographic distributions (age and gender) \\ \midrule
    Estimated audience size \\ \midrule
    Impressions \\ \midrule
    Amount spent (EUR) \\ \midrule
    Platform(s) on which the ad is published \\ 
    \bottomrule
\end{tabular}
\label{tab:ad-library-variables}
\end{table}

\clearpage
\section{Meta targeting categories}
\label{supp:targeting_categories}

\begin{figure}[H]
    \centering
    \includegraphics[width=.65\linewidth]{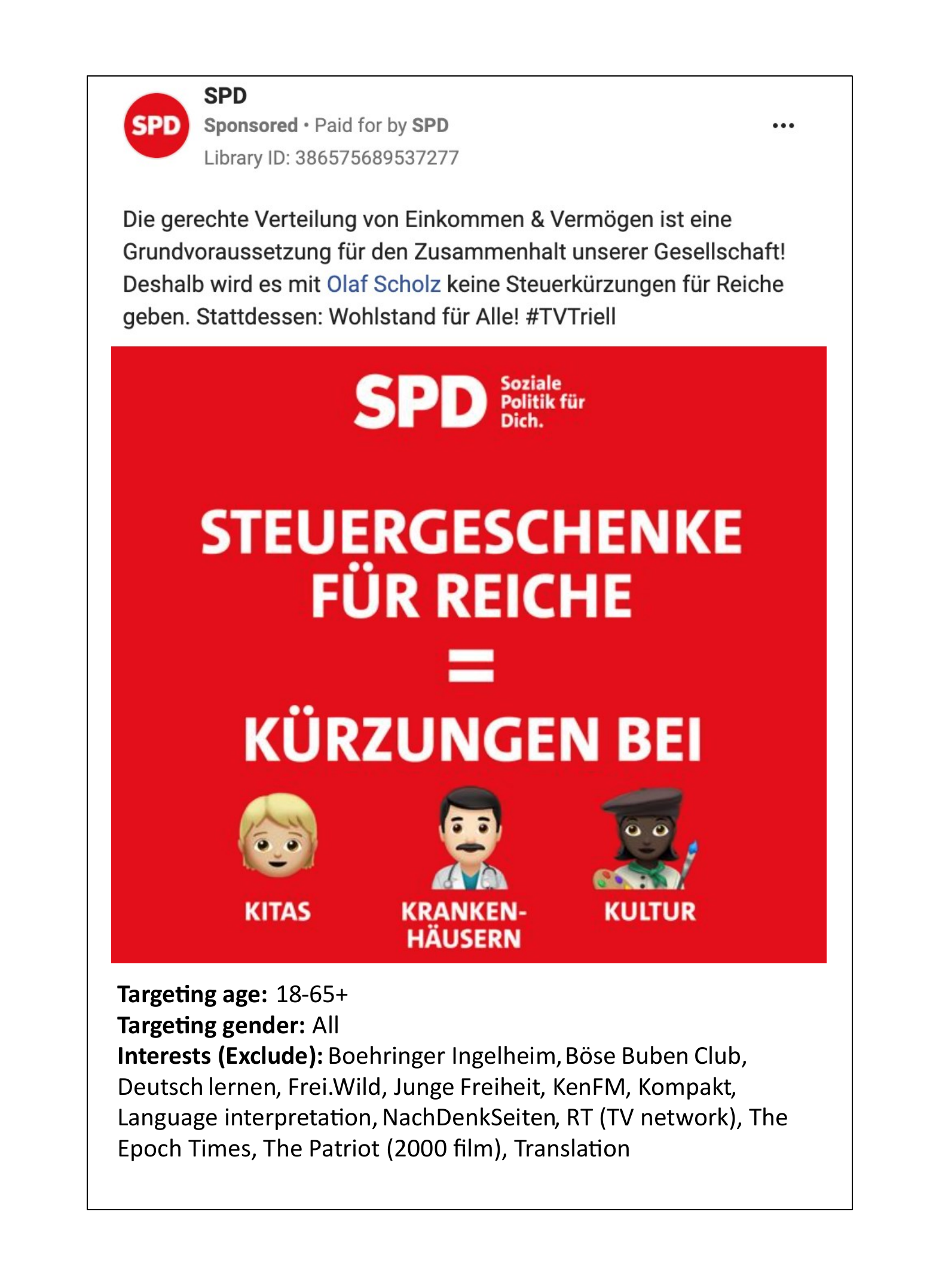}
    \caption{Example of an ad with targeting information as provided by Meta.}
    \label{fig:ad-targeting-example}
\end{figure}

Meta allows advertisers to include or exclude users based on a set of targeting categories to reach specific user groups on their platforms Facebook and Instagram. The full list of available targeting categories is shown in \Cref{tbl:targeting_categories}. For a more detailed description, we refer readers to \url{https://developers.facebook.com/docs/fort-ads-targeting-dataset/table-schema}. \Cref{fig:ad-targeting-example} provides an example ad with targeting information as provided by Meta.

\begin{table}[H]
    \centering
    \caption{Targeting categories available to advertisers on Meta. Advertisers can include or exclude users based on the targeting categories.}
    \label{tbl:targeting_categories}
    \begin{tabular}{c}
    \toprule
     Include/exclude interest \\ \midrule
     Include/exclude industry \\ \midrule
     Include/exclude parents \\ \midrule
     Include/exclude job title \\ \midrule
     Include/exclude employer \\ \midrule
     Include/exclude behavior \\ \midrule
     Include/exclude field of study \\ \midrule
     Include/exclude life event \\ \midrule
     Include/exclude school \\ \midrule
     Include/exclude education level \\ \midrule
     Include/exclude relationship status \\ \midrule
     Include/exclude income \\ \midrule
     Include/exclude undergrad years \\ \midrule
     Include/exclude custom audience \\ \midrule
     Include/exclude lookalike audience \\ \midrule
     Include/exclude connection \\ \midrule
     Include/exclude friend connection \\ 
     \bottomrule
    \end{tabular}
\end{table}

\clearpage
\section{Hyperparameter tuning}
\label{supp:hyperparameters}

For the training of the classifiers reported in the main paper, we use 10-fold cross-validation in combination with a grid search. In particular, we vary (1)~the number of trees used for a forest (\emph{$N$ estimators}), (2)~the number of variables to consider at each split (\emph{Max features}), (3)~maximum depth of the tree (\emph{Max depth}), (4)~minimum samples to split a node (\emph{Min node}), (5)~minimum samples in a leaf (\emph{Min leaf}), and (6)~whether to bootstrap samples when building trees (\emph{Bootstrap}). The corresponding tuning range for each parameter is shown in Supplementary~\Cref{tbl:hyper_rf}.

\begin{table}[!h]
    \caption{Tuning range of the grid search for random forest (RF) hyperparameters. $n$ refers to the total number of variables available for training.}
    \label{tbl:hyper_rf}
    \footnotesize
    \centering
    \begin{tabular}{ll}
        \toprule
        \textbf{Hyperparameter} & \textbf{Tuning Range} \\
        \midrule
        %\multirow{6}{*}{RF} & 
        \emph{$N$ estimators} & [100, 200, 300] \\
        \emph{Max features} & [$\sqrt{n}$, $\log_2 n $] \\
        \emph{Max depth} & [None, 10, 20, 30] \\
        \emph{Min node} & [2, 5, 10] \\
        \emph{Min leaf} & [1, 2, 4] \\
        \emph{Bootstrap} & [True, False] \\
        \bottomrule
    \end{tabular}
\end{table}

\clearpage
\section{Robustness checks}
\label{supp:robustness_checks}

We performed a series of checks to ensure the robustness of our results. In particular, we performed robustness checks regarding the (1)~machine learning model and (2)~platform heterogeneity.

First, we checked how our results change when using another machine learning classifier to predict impressions-per-EUR. In particular, we trained an XGBoost model using the same training procedure as outlined for the random forest model. However, the XGBoost model resulted in a lower prediction performance ($\mathit{RMSE}=$127.75 over 10 runs $\pm$ a s.d. of 3.91) when predicting impressions-per-EUR on the hold-out set. Nevertheless, our main findings remain consistent, i.e., (1)~our model can only explain a fraction of the variance in our data ($R^2=$0.36 $\pm$ a s.d. of 0.02 over 10 runs) and (2)~the far-right \emph{AfD} consistently achieves more impressions-per-EUR than predicted by our model. 

\begin{figure}[H]
    \centering
    \includegraphics[width=0.8\textwidth]{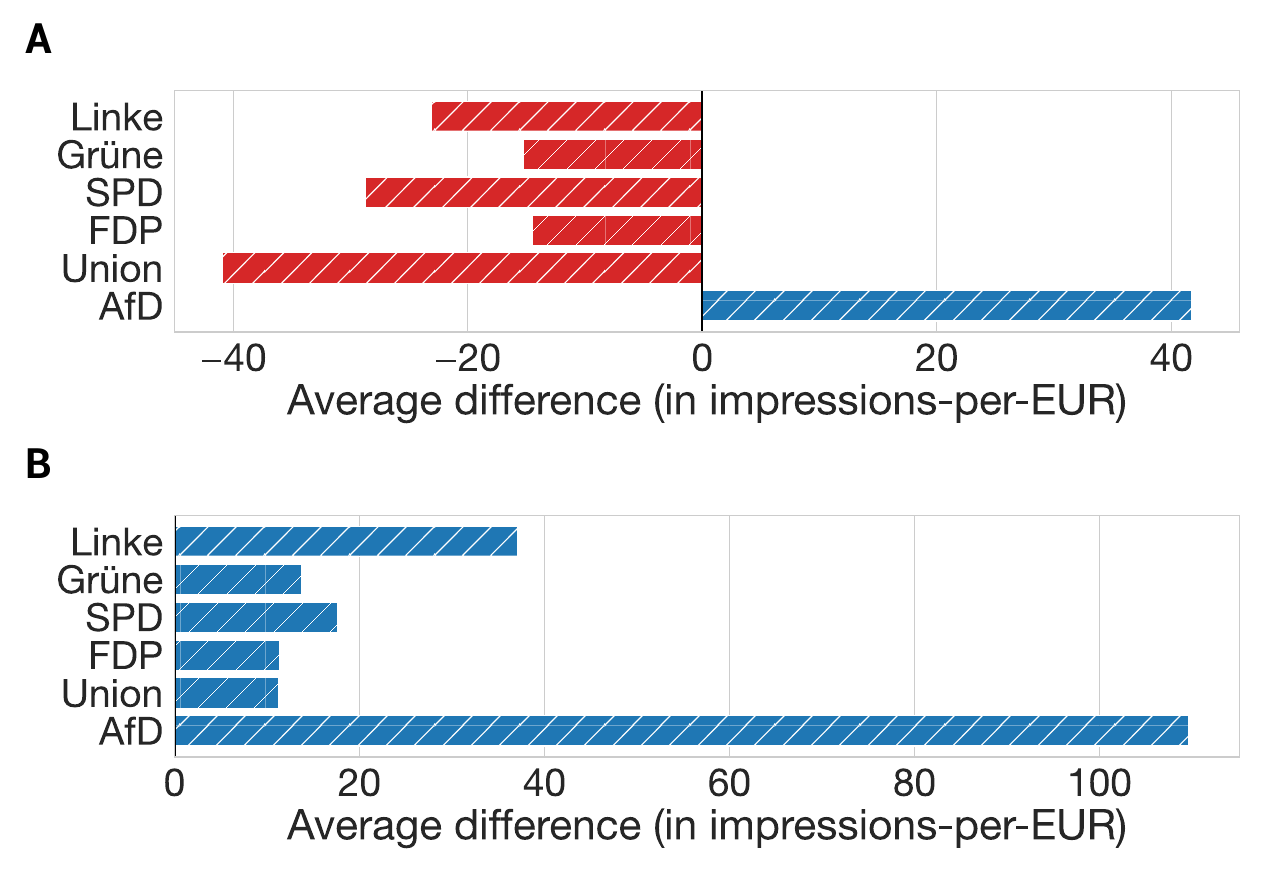}
    \caption{Average difference between actual vs. predicted impressions-per-EUR based on our machine learning model over 10 runs for political ads published on \figletter{a}, Facebook and \figletter{b}, Instagram.}
    \label{fig:robustness_check}
\end{figure}

Second, we checked the heterogeneity of our results with respect to the platform an ad was published. Specifically, we re-trained our random forest model but only used (1)~ads that were published on Facebook and (2)~ads that were published on Instagram. Hence, we trained two additional random forest model focusing on political ads from (1)~Facebook or (2)~Instagram. The results are shown in \Cref{fig:robustness_check}. Consistent with the results from our main analysis, we find systematic differences between actual impressions-per-EUR on the platform and predicted by our model. Importantly, we find that the far-right \emph{AfD} consistently achieves more impressions-per-EUR than predicted by our model.

\end{document}